\documentclass[fleqn,usenatbib]{mnras}

\usepackage{newtxtext,newtxmath}
\usepackage[T1]{fontenc}
\usepackage{ae,aecompl}

\usepackage{graphicx}	% Including figure files
\usepackage{amsmath}	% Advanced maths commands
\usepackage{amssymb}	% Extra maths symbols

\newcommand{\kms}{\mbox{km\,s$^{-1}$}}
\newcommand{\Msun}{\mbox{$M_{\odot}$}}
\newcommand{\Rsun}{\mbox{$R_{\odot}$}}

\newcommand{\Rjup}{\mbox{$R_{\rm Jup}$}}

\newcommand{\Rearth}{\mbox{$R_{\oplus}$}}

\newcommand{\ltsimeq}{\raisebox{-0.6ex}{$\,\stackrel
         {\raisebox{-.2ex}{$\textstyle <$}}{\sim}\,$}}

\title[K2-HERMES II. Complete results C1-C13]{K2-HERMES II. Planet-candidate properties from K2 Campaigns 1-13}

\author[R.A. Wittenmyer et al.]{
Robert A. Wittenmyer,$^{1}$\thanks{E-mail: rob.w@usq.edu.au (RW)}
Jake T. Clark,$^{1}$
Sanjib Sharma,$^{2}$
Dennis Stello,$^{3}$
\newauthor Jonathan Horner,$^{1}$
Stephen R. Kane,$^{4}$
Catherine P. Stevens,$^{5}$
Duncan J. Wright,$^{1}$
\newauthor Lorenzo Spina,$^{6}$
Klemen \v{C}otar,$^{14}$
Martin Asplund,$^{9}$
Joss Bland-Hawthorn,$^{2}$
\newauthor Sven Buder,$^{7,8,9}$
Andrew R. Casey,$^{6}$
Gayandhi M. De Silva,$^{2}$
Valentina D'Orazi,$^{10}$
\newauthor Ken Freeman,$^{9}$
Janez Kos,$^{2}$
Geraint Lewis,$^{2}$
Jane Lin,$^{2}$
Karin Lind,$^{8,11}$
\newauthor Sarah L. Martell,$^{3,7}$
Jeffrey D. Simpson,$^{3}$
Daniel B. Zucker,$^{12,13}$
Tomaz Zwitter$^{14}$
\\
$^{1}$University of Southern Queensland, Centre for Astrophysics, USQ Toowoomba, QLD 4350 Australia\\
$^{2}$Sydney Institute for Astronomy, School of Physics, University of Sydney, NSW 2006, Australia\\
$^{3}$School of Physics, University of New South Wales, Sydney 2052, Australia\\
$^{4}$Department of Earth and Planetary Sciences, University of California, Riverside, CA 92521, USA\\
$^{5}$Department of Physics, Westminster College, New Wilmington, PA 16172, USA\\
$^{6}$ Monash Centre for Astrophysics, School of Physics and Astronomy, Monash University, VIC 3800, Australia\\
$^{7}$ARC Centre of Excellence for All Sky Astrophysics in 3 Dimensions (ASTRO 3D), Canberra, ACT 2611, Australia\\
$^{8}$Max Planck Institute for Astronomy (MPIA), Koenigstuhl 17, 69117 Heidelberg, Germany\\
$^{9}$Research School of Astronomy \& Astrophysics, Australian National University, Canberra, ACT 2611, Australia\\
$^{10}$INAF Osservatorio Astronomico di Padova, vicolo dell'Osservatorio 5, 35122, Padova, Italy\\
$^{11}$Department of Physics and Astronomy, Uppsala University, Box 516, SE-751 20 Uppsala, Sweden\\
$^{12}$Department of Physics \& Astronomy, Macquarie University, Sydney, NSW 2109, Australia\\
$^{13}$Research Centre in Astronomy, Astrophysics \& Astrophotonics, Macquarie University, Sydney, NSW 2109, Australia\\
$^{14}$Faculty of Mathematics and Physics, University of Ljubljana, Jadranska 19, 1000 Ljubljana, Slovenia\\
}

\date{Accepted XXX. Received YYY; in original form ZZZ}

\pubyear{2020}

\begin{document}
\label{firstpage}
\pagerange{\pageref{firstpage}--\pageref{lastpage}}
\maketitle

% Abstract of the paper
\begin{abstract}
Accurate and precise radius estimates of transiting exoplanets are critical for understanding their compositions and formation mechanisms.  To know the planet, we must know the host star in as much detail as possible.  We present complete results for planet-candidate hosts from the K2-HERMES survey, which uses the HERMES multi-object spectrograph on the Anglo-Australian Telescope to obtain R$\sim$28,000 spectra for more than 30,000 K2 stars.  We present complete host-star parameters and planet-candidate radii for 224 K2 candidate planets from C1-C13.  Our results cast severe doubt on 30 K2 candidates, as we derive unphysically large radii, larger than 2\Rjup.  This work highlights the importance of obtaining accurate, precise, and self-consistent stellar parameters for ongoing large planet search programs - something that will only become more important in the coming years, as TESS begins to deliver its own harvest of exoplanets.
\end{abstract}

%We discuss the properties of the K2 planet sample as functions of age, metallicity, and other key stellar properties.  WE DIDN'T ACTUALLY DO THAT

% Select between one and six entries from the list of approved keywords.
% Don't make up new ones.

%%%%%%%%%%%%%%%%%%%%%%%%%%%%%%%%%%%%%%%%%%%%%%%%%%

%%%%%%%%%%%%%%%%% BODY OF PAPER %%%%%%%%%%%%%%%%%%

\section{Introduction} \label{sec:intro}

With the discovery of the first planets orbiting other stars \citep{GammaCeph,Latham,ps1257,51peg}, humanity entered the 'Exoplanet Era'.  For the first time, we had confirmation that the Solar system was not unique, and began to realise that planets are ubiquitous in the cosmos \citep[e.g.][]{fressin13,winn15,h19}.  At the same time, we learned that planetary systems are far more diverse than we had previously imagined.  We discovered planets denser than lead and more insubstantial than candy floss \citep{ross458c,Candy,dense,johns18}, found a myriad of systems containing giant planets orbiting perilously close to their host stars \citep[e.g.][]{51peg,Masset03,Bouchy05,HotJ2,HotJ1,albrecht12}, and discovered others with planets moving on highly elongated, eccentric orbits, similar to those of comets in the Solar system \citep[e.g.][]{Witt07,Tamuz08,Har15,HD76920}.  We even uncovered two types of planets that have no direct analogue in the Solar system -- the super-Earths and sub-Neptunes \citep[e.g.][]{SE1,SE3,SE2,SE4,SE5}.

The rate at which we found new exoplanets was boosted dramatically by the launch of the \textit{Kepler} spacecraft in 2009. In the years that followed, \textit{Kepler} performed the first great census of the Exoplanet Era.  In doing so, it revolutionised exoplanetary science, discovering some 2347 validated planets\footnote{as of 2020 Feb 26, from the NASA Exoplanet Archive, \url{https://exoplanetarchive.ipac.caltech.edu/}. A further 2420 candidate planets were found during the {\it{Kepler}} main mission, and still await confirmation.}, and finding hundreds of multiply-transiting systems \citep[e.g.][]{Borucki10,Batalha13,Mullaly15}.  After the failure of its second reaction wheel in 2013, the spacecraft was repurposed to carry out the ``K2'' mission \citep{howell14}.  \textit{Kepler's} golden years were spent in $\sim$80-day observations of fields along the ecliptic plane, with targets selected by the broader astronomical community for a wide range of astrophysical studies beyond planet search.  A total of 20 pointings (``campaigns'') were performed until the spacecraft station-keeping fuel was exhausted in 2018 October.  Altogether, the K2 mission observed more than 150,000 stars across 20 campaigns, resulting in 397 confirmed and 891 candidate planets to date\footnote{Planet data obtained from the NASA Exoplanet Archive, accessed 2020 Feb 26, at \url{https://exoplanetarchive.ipac.caltech.edu/}}. 

With the exception of the small number of directly imaged exoplanets \citep[e.g.][]{DI2,HR8799,HR87992,DI1}, our knowledge of the new worlds we discover has been gleaned indirectly.  We observe a star doing something unexpected, and infer the presence of a planet. Our knowledge of the planets we find in this manner is directly coupled to our understanding of their host stars.  For example, consider the case of a planet discovered using the transit technique. By measuring the degree to which the light of the planet's host star is attenuated during the transit, it is possible to infer the planet's size. The larger the planet, the more light it will block, and the greater the dimming of its host star.  As a result, it is relatively straightforward to determine the size of the planet \textit{relative to its host star}.  When converting those measurements to a true diameter for the newly discovered world, however, one must base that diameter on the calculated/assumed size of the host star.  Any uncertainty in the size of the host carries through to the determination of the size of the planet. 

For that reason, it is critically important for us to be able to accurately characterise the stars that host planets.  The more information we have about those stars, and the more precise those data, the more accurately we can determine the nature of their orbiting planets. 

Over the past few years, the Galactic Archaeology with HERMES survey (GALAH) has been gathering highly detailed spectra of a vast number of stars in the local Solar neighbourhood \citep[e.g.][]{desilva15,martell17,DR2}.  The survey uses the High Efficiency and Resolution Multi-Element Spectrograph (HERMES) on the Anglo-Australian Telescope \citep{HermesPaper,Hermes2} to simultaneously obtain approximately 400 spectra in a given exposure.  Analysis of those high-resolution spectra allows the determination of a variety of the properties of those stars, along with the calculation of accurate abundances for up to thirty different elements in their outer atmospheres.  GALAH aims to survey a million stars, facilitating an in-depth study of our Galaxy's star formation history - and has already yielded impressive results \citep[e.g.][]{Galah1,Galah2,Galah3,Galah6,Galah4,Galah5,Galah7,Galah8,Galah9}. Whilst the data obtained by the GALAH survey is clearly of great interest to stellar and Galactic astronomers, it can also provide information of critical importance to the exoplanet community.  For that reason, in this work we describe the results of the K2-HERMES survey, whose design follows that of the main GALAH program, but is designed specifically to maximise the scientific value of the plethora of exoplanets and oscillating stars discovered during \textit{Kepler's} K2 mission \citep{witt18,sharma19}. 

K2-HERMES is a survey born out of the urgent need for accurate, precise, and self-consistent physical parameters for stars including those hosting candidate planets.  Using the same instrumental setup and data processing pipelines as GALAH, the K2-HERMES survey aims to collect a spectrum for as many K2 target stars as possible in a given color-magnitude limited sample.  For each target so observed, we compute spectroscopic stellar parameters ($T_\mathrm{eff}$, log $g$, [Fe/H]), as well as the derived physical parameters such as mass, radius, luminosity, and age.  The HERMES instrument was specifically designed to measure the chemical abundances of up to thirty elements for the GALAH survey, and so those abundances are also delivered by the standard GALAH data processing pipeline.  A forthcoming paper, Clark et al. (2020, in prep), will present a detailed analysis of the chemical abundance results in the context of the \textit{Transiting Exoplanet Survey Satellite} mission, \textit{TESS}.  

In this paper, we present the complete results of planet-candidate properties from the K2-HERMES survey for K2 campaigns 1-13.  In Section 2, we briefly describe the observing strategy and data analysis procedures, and we detail how the stellar physical parameters have been derived.  Section 3 gives the physical properties of the K2 planet candidates and their host stars.  Finally, in Section 4, we place our results in context and present our conclusions.

%-----------------------------------------------------------------------
\section{Observations and Data Analysis} \label{sec:style}

%We use the High Efficiency and Resolution Multi-Element Spectrograph (HERMES), which can obtain spectra of up to 360 science targets simultaneously \citep{barden10,brzeski,heijmans12,sheinis15}.  

\begin{figure}
\includegraphics[width=\columnwidth]{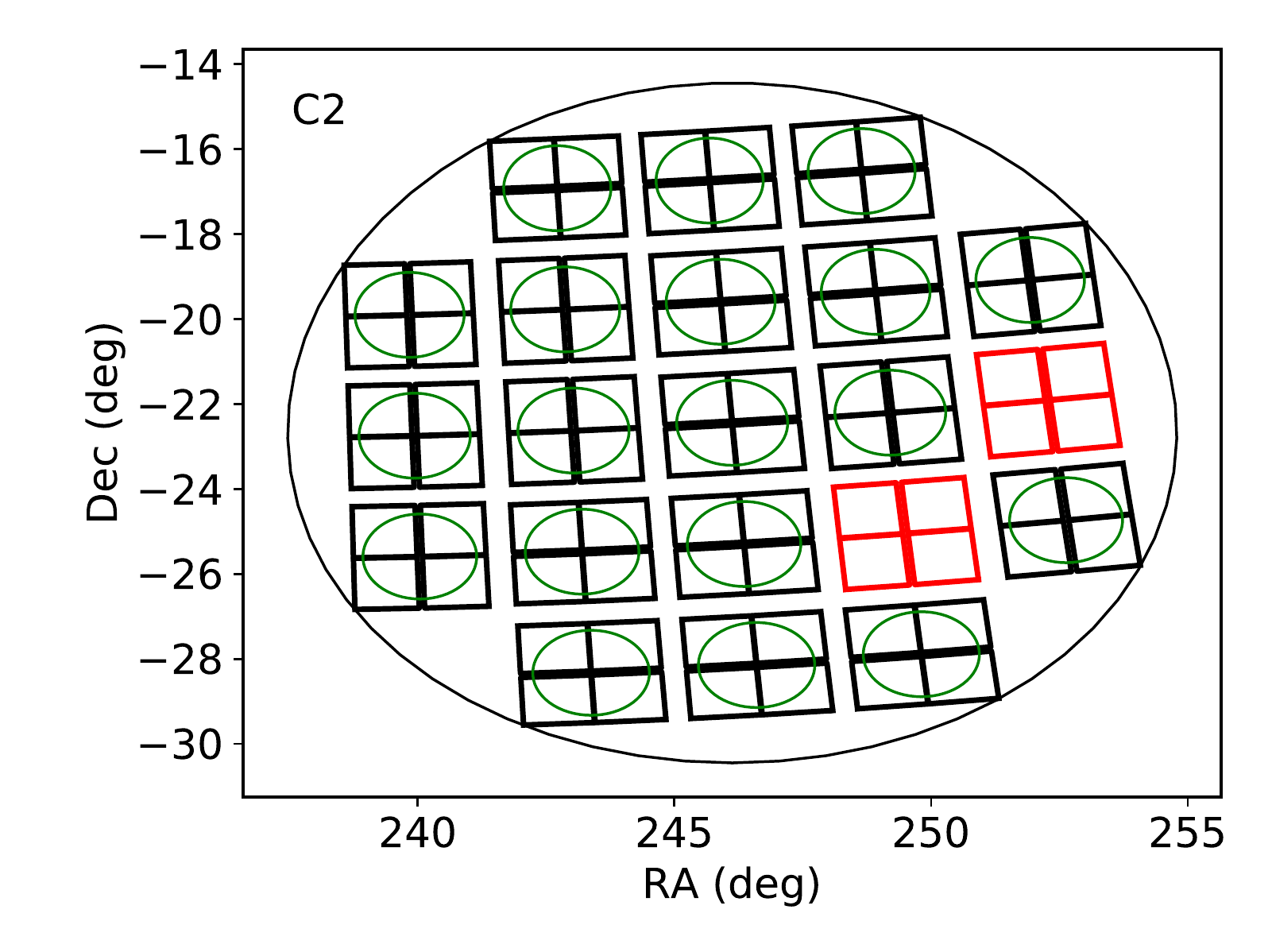}
\caption{The \textit{Kepler} field of view and the layout of its CCD modules, overlaid with the HERMES field of view (green circles).  The red modules are inoperative. }
\label{fig:fov}
\end{figure}

Target selection for the K2-HERMES program is described fully in our previous work \citep{witt18,sharma19}.  Figure~\ref{fig:fov} shows the HERMES field of view overlaid on the \textit{Kepler} field.  For this study, we selected all K2 planet candidate host stars which had been observed in the K2-HERMES program.

%is essentially unbiased, since the star densities in K2 ecliptic fields are well-matched to the 360 science fibres available for each 1-degree (radius) observing field.  To prevent excessive cross-talk between fibres, a given field is observed twice, as a ``bright'' ($10<V<13$) and ``faint'' ($13<V<15$) exposure \citep[as described in][]{martell17}. Each field contains an average of 210 K2 targets, so the end result is that every K2 target is observed, 
%[REALLY? EVERY TARGET? OR DO YOU MEAN EVERY TARGET IN THE COLOUR-MAG RANGE THAT WE CHOOSE]  This is a tautology: "we observe every target that we observe."
% except those that fall near the corners of the K2 CCD modules (Figure~\ref{fig:fov}).    

% raw reduction paragraph copied directly from first K2 paper.
%The raw reduction procedure, described fully in \citet{kos17} and \citet{sharma17}, is in brief: (1) raw reduction is performed with a custom IRAF-based pipeline, (2) four basic parameters ($T_\mathrm{eff}$, log $g$, [Fe/H], and radial velocity) and continuum normalisation are calculated with a custom pipeline ``GUESS'' by matching the observed normalized spectra to synthetic templates.  A grid of AMBRE synthetic spectra is used for this purpose \citep{delaverny12}.

% Kepler FOV fig stolen from first K2 paper.

\subsection{Determination of stellar parameters}

We find 199 stars hosting 224 K2 planet candidates for which K2-HERMES spectra are available.  The reduction and analysis procedures are identical to those of the GALAH and TESS-HERMES surveys, as described fully in \citep{kos17,DR2,sharma17}.

% this paragraph was copied directly from first K2 paper. and some from Sharma+2019. Referee called me on it so it is now blue.

%The spectroscopic stellar parameters have been estimated with a combination of classical spectrum synthesis for a representative reference set of stars and a data-driven approach to propagate the high-fidelity parameter information with higher precision onto all the stars in the K2-HERMES survey.  The method is identical to that used by the TESS-HERMES survey \citep{sharma17}, and is briefly outlined as follows.  The reduction is done using a custom IRAF-based pipeline \citep{kos17}.  The spectroscopic analysis is done using the GALAH pipeline and is described in \citet{DR2}.  It uses Spectroscopy Made Easy (SME) to first build a training set by means of a model driven scheme \citep{piskunov17}.  Next, \textit{The Cannon} \citep{ness15} is used to estimate the stellar parameters and abundances by means of a data driven scheme.

%First, we use the spectrum synthesis code Spectroscopy Made Easy (SME) by \citet{piskunov17} to analyse the reference set.  This training set includes samples of stars with external parameter estimates, \textit{Gaia} benchmark FGK stars, and stars with asteroseismic information from K2 Campaign 1 \citep{stello17}.  Next, we use these SME results as input labels for the training set for \textit{The Cannon} \citep{ness15} to propagate the analysis to all stars.  This procedure is identical to that described in the GALAH second data release \citep{DR2}.\\

 With a self-consistent set of spectroscopic parameters in hand ($T_\mathrm{eff}$, log $g$, [Fe/H]), we derived the stellar physical parameters using the \texttt{isochrones} Python package \citep{morton15}.  \texttt{isochrones} is a Bayesian isochronic modeller that determines the mass, radius and age of stars given various photometric and spectroscopic inputs using MESA Isochrones \& Stellar Tracks (MIST) \citep{MIST} grids.  For our analysis, we used the effective temperature ($T_\mathrm{eff}$), surface gravity ($\log{g}$), 2MASS ($H$, $J$, $K_s$) \citep{twomass} and \textit{Gaia} ($G$, $G_{RP}$, $G_{BP}$) photometric magnitudes along with parallax values obtained by \textit{Gaia} DR2 \citep{gaia18} where available. 

Accurate isochrone models rely upon a star's global metallicity, commonly referred as [M/H]. The assumption that the iron abundance [Fe/H] can be a proxy (or even equal) to [M/H] breaks down for metal-poor stars. In these metal-poor stars, the radiative opacity can be heavily affected by alpha-elements, in our case Mg, Si, Ca and Ti. Including alpha-elements into our global metallicity thus better predicts the physical parameters derived with \texttt{isochrones}.  We calculate our [$\alpha$/Fe] values through equation \ref{eq:alpha}, which is the exact procedure taken by GALAH DR2:

\begin{equation}\label{eq:alpha}
[\alpha/Fe] = \frac{\sum{\frac{[X/Fe]}{(e\_[X/Fe])^2}}}{\sum{(e\_[X/Fe])}^{-2}}
\end{equation}

\noindent where X = Mg, Si, Ca, Ti and e\_[X/Fe] is the abundance's associated error. [$\alpha$/Fe] is calculated even if one or more of these elements are missing. From [Fe/H] and [$\alpha$/Fe], we can then calculate [M/H] through a relationship between these quantities laid out in \citet{alphaMH}:

\begin{equation}
[M/H] = [Fe/H] + \log_{10}{\big(0.638*f_\alpha + 0.362\big)}
\end{equation}

\noindent where $f_\alpha$ is the $\alpha$-element enhancement factor given by $f_\alpha = 10^{[\frac{\alpha}{Fe}]}$. Our calculated [M/H] value is then used for our isochrone star model on top of the discussed parameters above.  After the model reaches convergence, median output values of the stellar mass, radius, density, age, bolometric luminosity and equivalent evolution phase and their corresponding 1-$\sigma$ errors are calculated from the posterior distributions.  We calculate the stellar luminosity by:
\begin{equation}
    \bigg(\frac{L}{L_\odot}\bigg) = \bigg({\frac{R}{R_\odot}}\bigg)^2\bigg({\frac{T}{T_\odot}}\bigg)^4
\end{equation}

\begin{figure}
\includegraphics[width=\columnwidth]{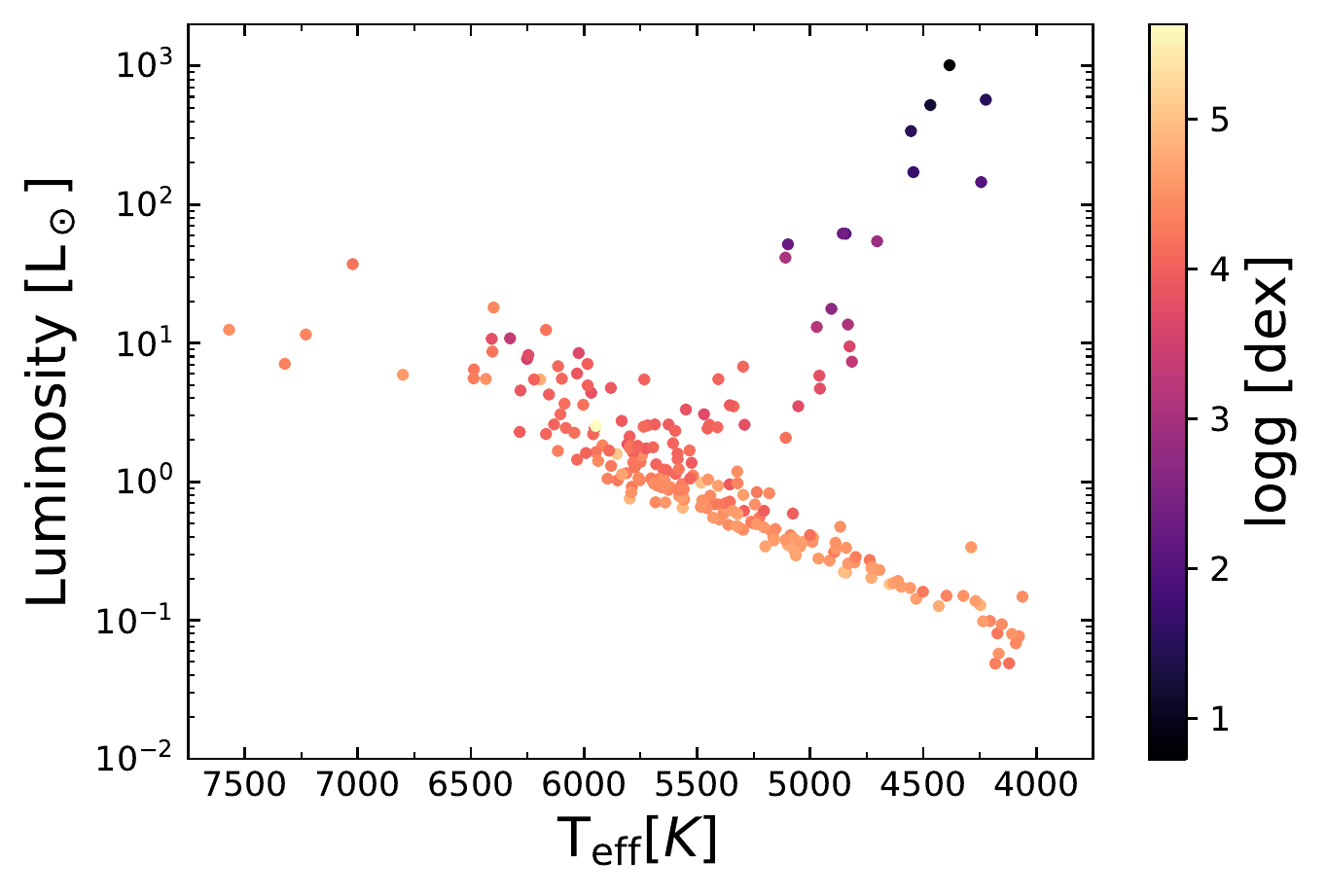}
\caption{H-R diagram of our K2-HERMES and \texttt{isochrones}-derived results for 199 K2 stars. }
\label{hrdiagram}
\end{figure}

A Hertzsprung-Russell diagram of our results is shown in Figure~\ref{hrdiagram}, based on our $T_\mathrm{eff}$, $\log{g}$, and \texttt{isochrones}-derived stellar luminosity.  This sanity check confirms that none of our 199 K2 stars fall in unphysical regions of parameter space.  Three stars, (EPIC\,201516974, 211351816, 211390903) show asteroseismic detections of the large frequency separation, $\Delta \nu$, and the frequency at maximum power, $\nu_{\mathrm{max}}$. For these detections we used EVEREST K2 light curves \citep{luger16} that we analysed following the approach by \citet{stello17}, which uses the method by \citet{huber09} with the improvements described in \citet{huber11} and in \citet{yu18}.  Then, using the seismic $\Delta \nu$ and $\nu_{\mathrm{max}}$ and the methods of \citet{hon18} and \citet{sharma16}, we derived physical parameters for these three stars and give them in Table~\ref{tab:seismology} alongside our spectroscopic results from \texttt{isochrones}.

% seismology results mini table here

\begin{table*}
    \centering
    \begin{tabular}{lllllll}
    \hline
    EPIC      &   log $g$  & Radius (\Rsun)  &  Mass (\Msun) &  log $g$   &  Radius (\Rsun)  &  Mass (\Msun) \\
       & Seismology  & & & K2-HERMES & & \\
    \hline
201516974 &  2.934$\pm$0.010  & 5.26$\pm$0.16 & 0.87$\pm$0.07 & 2.69$\pm$0.16 & 5.84$\pm$0.25 & 1.30$\pm$0.14 \\
211351816 &  3.245$\pm$0.007 & 4.11$\pm$0.07 & 1.08$\pm$0.05 & 4.16$\pm$0.17 & 4.42$\pm$0.24 & 1.56$\pm$0.15 \\
211390903 &  2.626$\pm$0.022  & 8.8$\pm$0.5 & 1.19$\pm$0.20 & 2.89$\pm$0.19 & 11.10$\pm$0.56 & 1.73$\pm$0.28 \\
\hline
    \end{tabular}
    \caption{Stellar parameters derived from seismology, and comparison with the spectroscopic results from K2-HERMES. }
    \label{tab:seismology}
\end{table*}

%epic, evstate, numax, e_numax, dnu, e_dnu, sradius, e_sradius, mass, e_mass, logg_seism, e_logg_seism, radius_rob, e_radius_rob, mass_rob, e_mass_rob
%201516974, 1.0, 104.963, 2.388, 10.149, 0.105, 5.257898439199262, 0.16422724957383475, 0.8654133333126983, 0.07051686619138296, 2.933559570439109, 0.010163968448486618, 5.84, 0.254, 1.3, 0.137
%211351816, 1.0, 216.859, 3.161, 16.854, 0.074, 4.110413445771231, 0.07435175597334384, 1.083758362263325, 0.05482608618512758, 3.2451218209445782, 0.00686974462541594, 4.42, 0.243, 1.56, 0.153
%211390903, 2.0, 52.769, 2.672, 5.661, 0.085, 8.790699401830658, 0.5213343011584325, 1.1908425193734997, 0.19619997388154028, 2.6257678482335507, 0.022215490275242205, 11.1, 0.563, 1.73, 0.283

\begin{figure}
\includegraphics[width=\columnwidth]{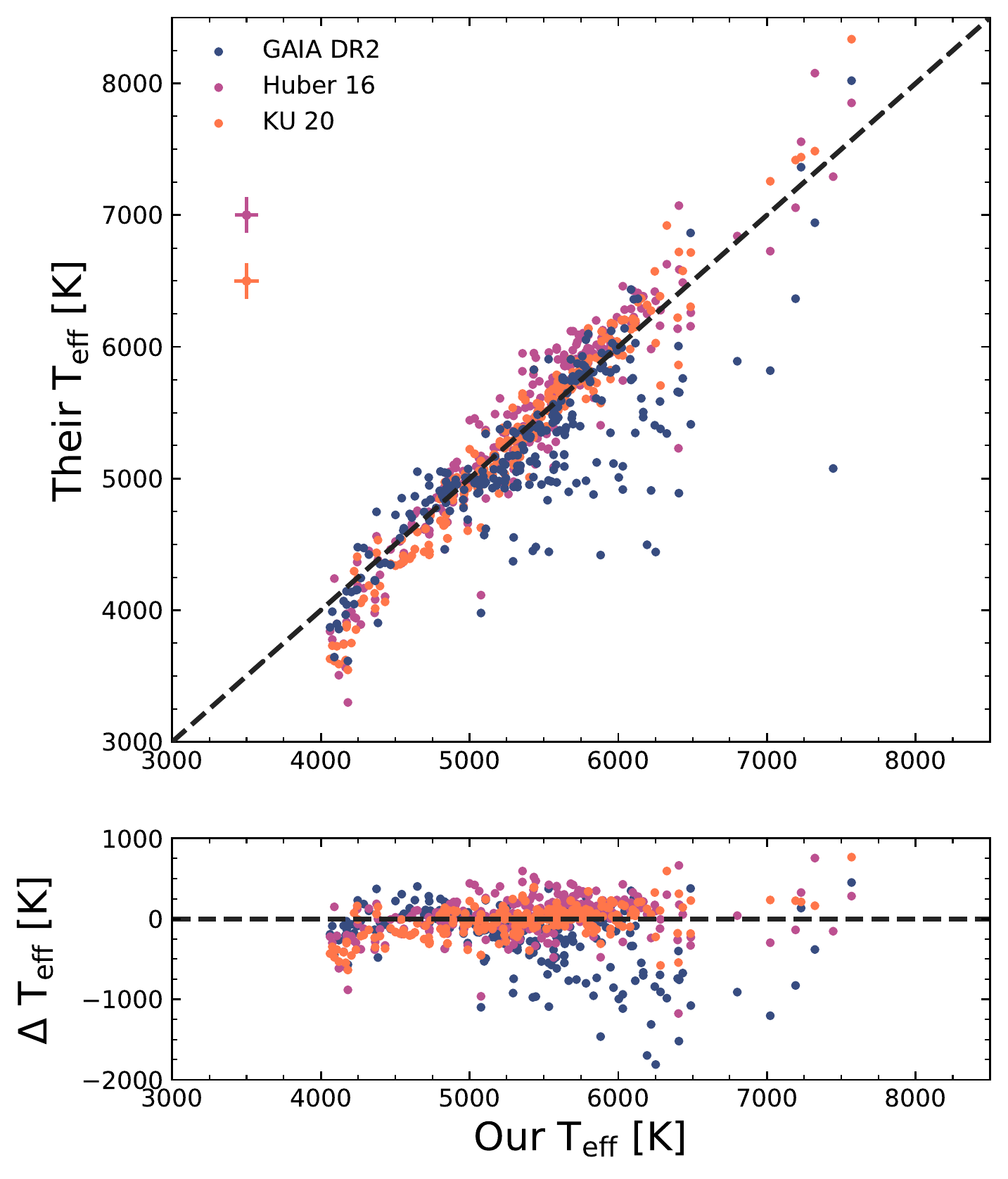}
\caption{Comparison of our revised T$_{eff}$ with published values.  The RMS differences are: Gaia -- 29\,K, H16 -- 16\,K, KU20 -- 11\,K.  Median error bars are also shown. }
\label{teffcomp}
\end{figure}

\begin{figure}
\includegraphics[width=\columnwidth]{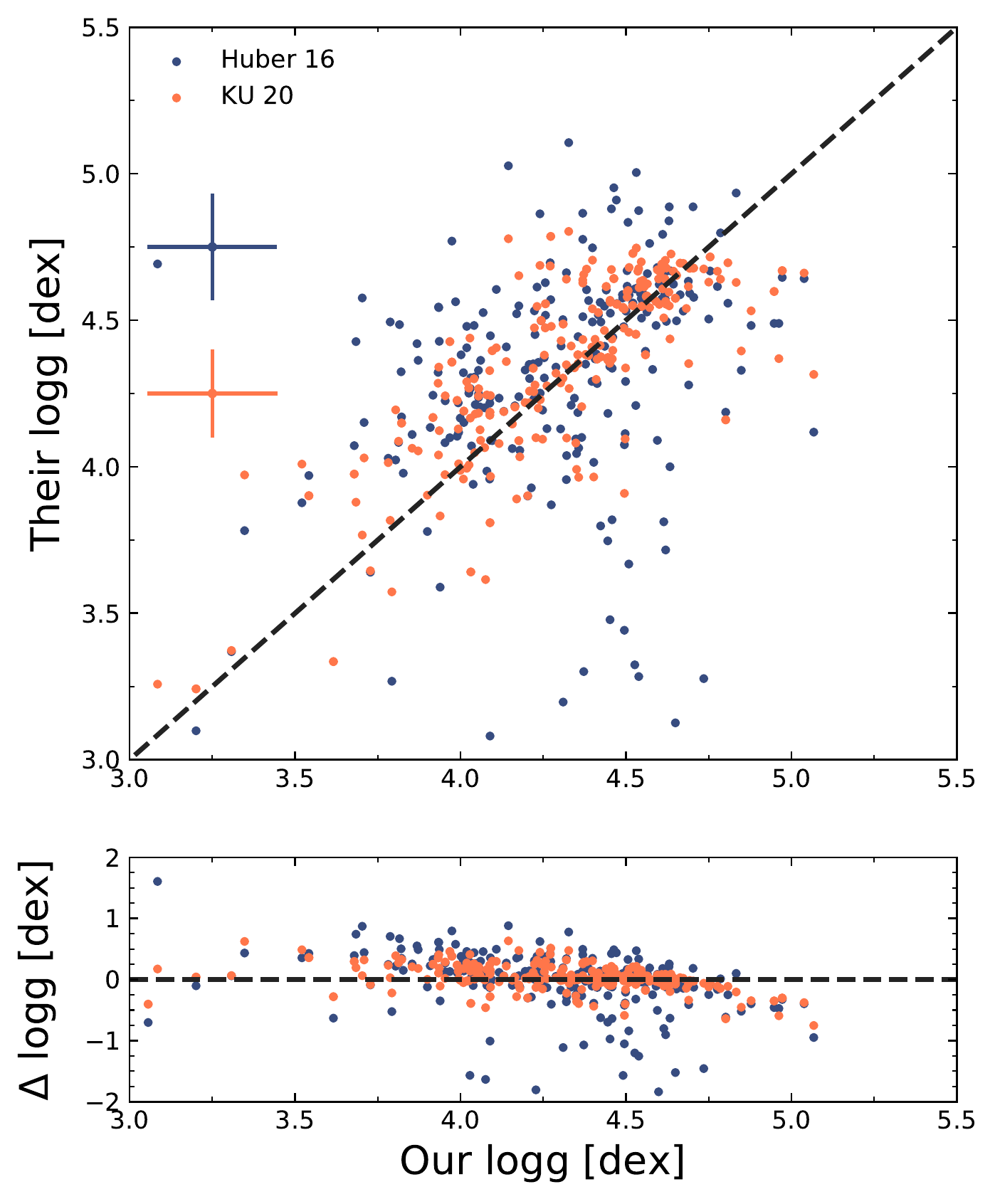}
\caption{Comparison of our revised log $g$ with published values.  The RMS differences are: H16 -- 0.03 dex, KU20 -- 0.01 dex.  Median error bars are also shown. }
\label{loggcomp}
\end{figure}

\begin{figure}
\includegraphics[width=\columnwidth]{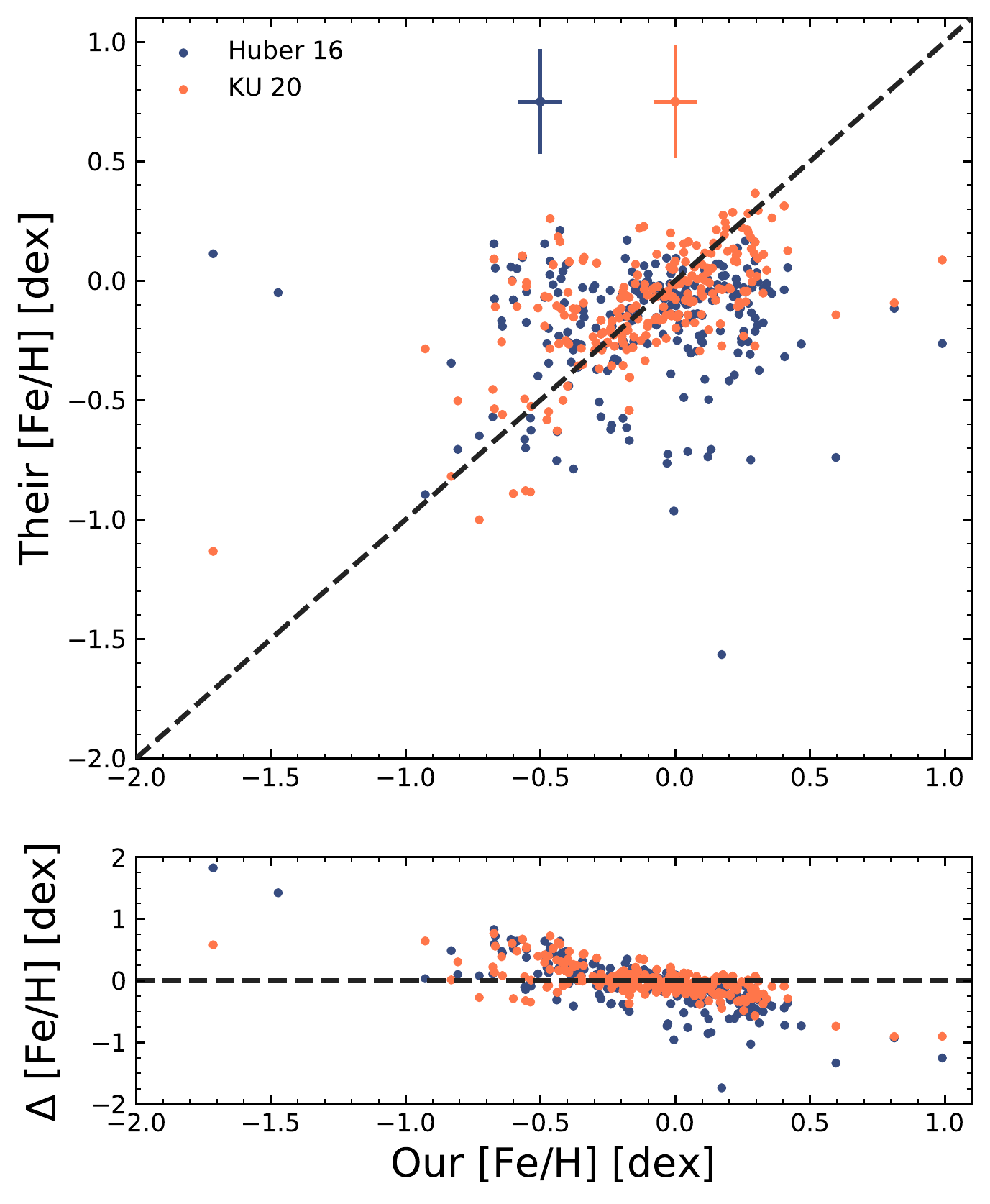}
\caption{Comparison of our revised [Fe/H] with published values.  The RMS differences are: H16 -- 0.02 dex, KU20 -- 0.01 dex.  Median error bars are also shown. }
\label{fehcomp}
\end{figure}

\begin{figure*}
\includegraphics[width=\columnwidth]{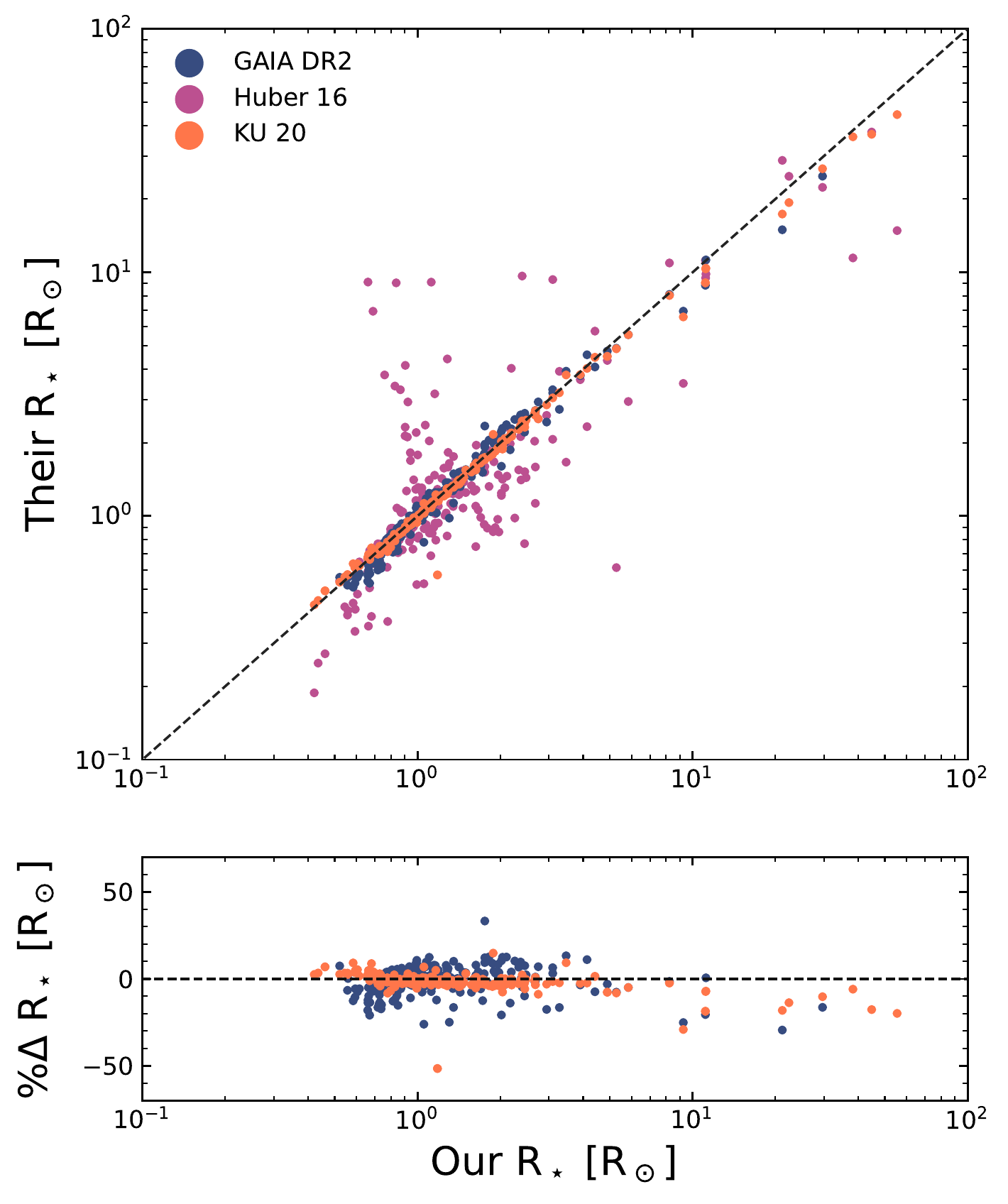}
\includegraphics[width=0.967\columnwidth]{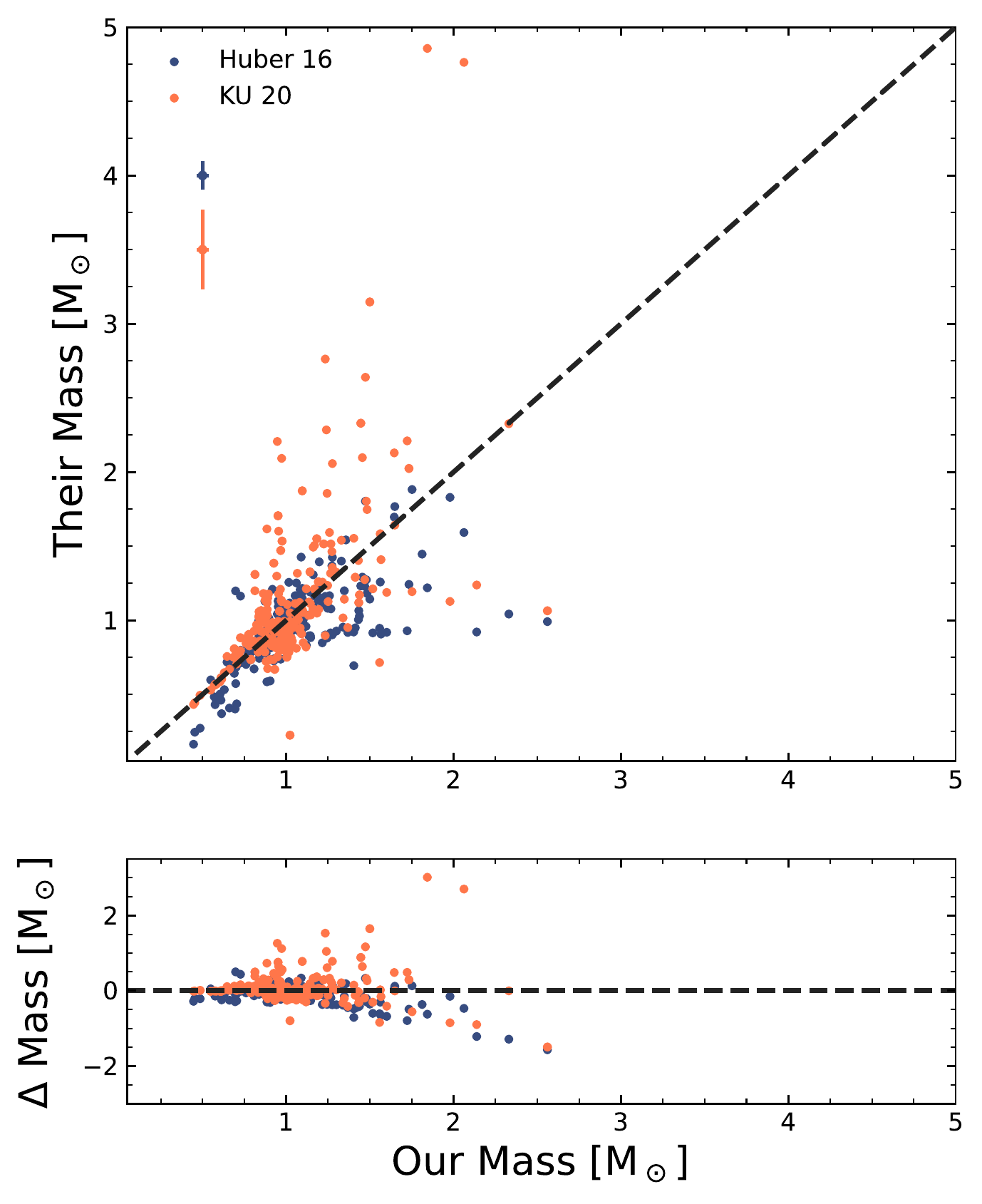}
\caption{Comparison of our derived stellar physical parameters with published values. }
\label{starcomps2}
\end{figure*}

The resulting stellar parameters are given in Table~\ref{tab:stellarparams}.  Our K2-HERMES results have the following median uncertainties: $T_\mathrm{eff}$: 74\,K, log\,$g$: 0.19 dex, [Fe/H]: 0.08 dex, $M_*$: 0.036\,\Msun, $R_*$: 0.019\,\Rsun.  Figures~\ref{teffcomp}--\ref{fehcomp} compare our K2-HERMES spectroscopic parameters with those presented by \citet{h16} (based largely on multicolour photometry), and recent results from \citet{hu20} based on LAMOST spectra.  Figure~\ref{starcomps2} shows the comparison between our derived stellar radii and masses and those of \citet{h16} and \citet{hu20}, as well as the radii inferred from \textit{Gaia}.    

A primary motivation for refining stellar parameters is to determine which planets would be best suited for follow-up activities \citep{chandler2016,kempton2018,ostberg2019}. This is particularly true of studies related to potentially habitable planets and the effect of stellar properties on the extent of the Habitable Zone (HZ) \citep{kane2014a,kane2018}.
The stellar parameters derived above were used to estimate several key properties of the known planets and their systems, shown in Table~\ref{tab:insolation}. We calculated the incident flux received by the planet in units of the solar constant ($F_\oplus$) using the semi-major axis and stellar luminosity. We further calculated the equilibrium temperature for each planet ($T_\mathrm{eq}$) using both "hot dayside" and well-mixed models, which assume that the planet re-radiates as a blackbody over $2\pi$ and $4\pi$ steradians respectively \citep{kane2011}. Finally, we calculated the HZ boundaries for each of the stars, using the formalism described by \citet{kopparapu2013,kopparapu2014}. We calculated the "runaway greenhouse" and "maximum greenhouse" boundaries (referred to as the "conservative" HZ) and the empirically derived "recent Venus" and "early Mars" boundaries (referred to as the "optimistic" HZ). A thorough description of these boundaries and how they are used is provided by \citet{kane2016}. Although all of the planets whose insolation properties are shown in Table~\ref{tab:insolation} are interior to the HZ, some of the planets do lie in the Venus Zone (VZ) \citep{kane2014b}. Terrestrial planets that lie within the VZ are also valued targets for follow-up activities as they can provide insight into the boundaries of habitability and the divergence of the Venus/Earth atmospheric evolution \citep{kane2019}. Further investigations of these systems may yet reveal additional planets within the HZ of the stars, increasing the value of those systems through comparative planetology studies of planets throughout the system.

%\citet{h16} (hereafter H16) presented a catalog of stellar parameters for 138,600 stars in the K2 Ecliptic Plane Input catalog (EPIC) for Campaigns 1-8.  Figure~\ref{starcomps1} shows a comparison of stellar spectroscopic parameters ($T_{eff}$, log $g$, [Fe/H]) obtained by K2-HERMES with those given in H16.  We find NN matches between the H16 and K2-HERMES catalogs.  Only stars hosting confirmed planets are shown here for clarity.  \textbf{need the numbers here: want median differences between us an H16 for the 3 sets of things}.  For $T_{eff}$ and log\,$g$, our results are consistent with H16, with a few outliers at $T_{eff}\sim$4000\,K, which is expected for the coolest dwarfs in our sample, owing to a paucity of similar stars in the training set \citep{sharma17,bensby14,torres12}.  In log\,$g$, we find 8 stars for which our value is more than $3\sigma$ discrepant from H16.  Nearly all of these stars were classified by H16 as giants, but we find them to be dwarfs.   

% this was figure 3 but is now 3 separate ones to embiggen the plots.

%\includegraphics[width=0.66\columnwidth]{logg.pdf}
%\includegraphics[width=0.66\columnwidth]{feh.pdf}
%\caption{Comparison of our revised spectroscopic stellar parameters with published values. }
%\label{starcomps1}
%\end{figure}

% this was figure 4 in submitted version.

%\clearpage

% ONLINE ONLY TABLE, first 10 rows shown here.  "stellarparamstable_nonlatex.txt"
% numbers are now correct and in EPIC order.
% Updated 2020 Feb 10 with new iso results. 
\begin{table*}
    \centering
    \begin{tabular}{llllll}
    \hline
    EPIC      &     T$_\mathrm{eff}$   &   log $g$   &  [Fe/H]   &  Mass (\Msun)  & Radius (\Rsun) \\
    \hline
201110617 &  4247.7 $\pm$  465.7 & 4.83 $\pm$  0.23 & -0.17 $\pm$  0.10 & 0.695 $\pm$ 0.020 &  0.663 $\pm$  0.009 \\
201127519 &  4737.0 $\pm$   58.1 & 4.23 $\pm$  0.17 &  0.15 $\pm$  0.07 & 0.832 $\pm$ 0.026 &  0.777 $\pm$  0.008 \\
201128338 &  4205.2 $\pm$   81.0 & 4.37 $\pm$  0.18 & -0.47 $\pm$  0.07 & 0.610 $\pm$ 0.012 &  0.594 $\pm$  0.007 \\
201132684 &  5407.0 $\pm$   54.8 & 4.37 $\pm$  0.17 &  0.10 $\pm$  0.07 & 0.915 $\pm$ 0.029 &  0.947 $\pm$  0.013 \\
201155177 &  4694.2 $\pm$   98.1 & 4.56 $\pm$  0.21 & -0.20 $\pm$  0.09 & 0.760 $\pm$ 0.025 &  0.727 $\pm$  0.014 \\
201160662 &  6486.5 $\pm$   68.9 & 4.25 $\pm$  0.19 & -0.81 $\pm$  0.08 & 1.240 $\pm$ 0.072 &  2.020 $\pm$  0.067 \\
201264302 &  4181.5 $\pm$  207.5 & 4.33 $\pm$  0.21 & -0.48 $\pm$  0.09 & 0.446 $\pm$ 0.025 &  0.421 $\pm$  0.006 \\
201390927 &  4288.2 $\pm$   71.9 & 4.57 $\pm$  0.19 & -0.30 $\pm$  0.08 & 0.884 $\pm$ 0.053 &  1.050 $\pm$  0.091 \\
201393098 &  5625.9 $\pm$   73.6 & 3.94 $\pm$  0.19 & -0.34 $\pm$  0.08 & 1.070 $\pm$ 0.039 &  1.700 $\pm$  0.040 \\
201403446 &  6132.3 $\pm$   59.9 & 4.05 $\pm$  0.18 & -0.47 $\pm$  0.07 & 1.060 $\pm$ 0.040 &  1.430 $\pm$  0.034 \\
\hline
    \end{tabular}
    \caption{Spectroscopic and derived stellar parameters.  The full version of this table is available online.}
    \label{tab:stellarparams}
\end{table*}

%these are the new ones! Added 7 stars. But not.
% 210559259 &  4914.6$\pm$ 187.5 & 4.56$\pm$0.22 & -0.03$\pm$0.09 & 0.761$^{+0.024}_{-0.026}$  &  0.710$^{+0.010}_{-0.009}$  \\ Zink09
% 211351097 &  5875.4$\pm$  74.1 & 4.33$\pm$0.20 &  0.06$\pm$0.08 & 1.055$^{+0.040}_{-0.038}$  &  1.132$^{+0.023}_{-0.023}$  \\ K2scicon poster "Mind the Gap" Dholakia at Berkeley
% 211821192 &  5766.9$\pm$ 176.3 & 4.57$\pm$0.21 & -0.19$\pm$0.09 & 0.945$^{+0.059}_{-0.052}$  &  1.021$^{+0.023}_{-0.023}$  \\ K2scicon poster "Mind the Gap" 
% 211953574 &  5679.1$\pm$  72.8 & 3.95$\pm$0.17 & -0.03$\pm$0.07 & 0.962$^{+0.034}_{-0.034}$  &  1.135$^{+0.016}_{-0.016}$  \\ K2scicon poster "Mind the Gap" 
% 220400100 &  4831.8$\pm$  83.7 & 4.54$\pm$0.21 & -0.11$\pm$0.09 & 0.753$^{+0.024}_{-0.025}$  &  0.716$^{+0.010}_{-0.009}$  \\ Zink09
% 247047370 &  5000.7$\pm$  70.5 & 4.18$\pm$0.18 &  0.27$\pm$0.07 & 0.875$^{+0.027}_{-0.019}$  &  0.863$^{+0.011}_{-0.010}$  \\ Zink09
% 247063356 &  6030.7$\pm$  68.2 & 4.09$\pm$0.19 & -0.07$\pm$0.08 & 1.051$^{+0.039}_{-0.039}$  &  1.088$^{+0.021}_{-0.021}$  \\ Zink09

% ANOTHER ONLINE ONLY TABLE "habzonetable_nonlatex.txt"
% Update 2020 Feb 19 with vetted list, new AU values from new Mstars.

\begin{table*}
    \centering
    \begin{tabular}{llllllll}
    \hline
    EPIC  &  Incident Flux  & T$_\mathrm{eq}$ (K) & T$_\mathrm{eq}$ (K) & HZ (au) & HZ (au) & HZ (au) & HZ (au) \\
         &  $F_{\oplus}$  &  hot dayside    &  well-mixed  &  inner, opt   &  inner, conserv  & outer, conserv  & outer opt \\
    \hline
201110617.01  &      566.1  &   1616.0  &   1358.9  &   0.29  &   0.37  &   0.69  &   0.72 \\
201127519.01  &       70.9  &    961.2  &    808.3  &   0.41  &   0.52  &   0.96  &   1.01 \\
201128338.01  &        3.4  &    451.2  &    379.4  &   0.25  &   0.32  &   0.60  &   0.64 \\
201132684.01  &      178.6  &   1211.0  &   1018.3  &   0.64  &   0.81  &   1.44  &   1.51 \\
201132684.02  &       87.7  &   1013.8  &    852.5  &   0.64  &   0.81  &   1.44  &   1.51 \\
201155177.01  &       57.3  &    911.4  &    766.4  &   0.38  &   0.48  &   0.88  &   0.93 \\
201160662.01  &     8243.2  &   3156.7  &   2654.4  &   1.83  &   2.31  &   4.02  &   4.24 \\
201264302.01  &     1722.6  &   2134.3  &   1794.7  &   0.18  &   0.23  &   0.42  &   0.45 \\
201390927.01  &      262.3  &   1333.2  &   1121.1  &   0.47  &   0.59  &   1.11  &   1.17 \\
201393098.01  &       73.7  &    970.7  &    816.3  &   1.22  &   1.54  &   2.73  &   2.88 \\
   \hline
    \end{tabular}
    \caption{Planetary insolation and Habitable Zone boundaries. The full version of this table is available online.}
    \label{tab:insolation}
\end{table*}

%-----------------------------------------------------------------------
\section{Planet Candidate Parameters}

Table~\ref{tab:planetparams} gives the properties of the 224 planet candidates from C1-C13 for which the K2-HERMES program has obtained spectra of their host stars.  The orbital period and relative radius $R_p/R_*$ are obtained from the NASA Exoplanet Archive, with the relevant references cited in Table~\ref{tab:planetparams}.  Where multiple published values exist, the most recent reference was chosen for our analysis.  The semimajor axis values have been recalculated based on the orbital period and the revised stellar masses given in Table~\ref{tab:stellarparams}.  We derived the planet-candidate radii by multiplying $R_p/R_*$ by the stellar radii obtained by \texttt{isochrones} as described above.  Uncertainties in the planetary radii result from the propagated uncertainties in $R_*$ and $R_p/R_*$.  As in our previous work \citep{witt18}, for those planet candidates without published uncertainties in $R_p/R_*$, we adopted the median fractional uncertainty of 0.0025 derived from the catalog of \citet{c16}.  

% 2020 Feb 14: updated radii here based on latest iso results. DONE
% 2020 Feb 14: AU are updated. DONE
% Added planets have been added (period and Rp/Rstar). DONE
% MNRAS can never get long tables to work!!
\begin{table*}
    \centering
    \begin{tabular}{llclllr}
    \hline
EPIC  &  K2 ID   &  Reference &  P (days)   &   a (au)  &   $R_p/R_*$    &   $R_p$ (\Rearth) \\ 
    \hline
   201110617  & K2-156      & 1 & 0.813149$\pm$0.000050  &  0.01510$\pm$0.00014 & 0.017041$\pm$0.0014  & 1.23$\pm$0.10     \\ 
   201127519  &             & 1 & 6.178369$\pm$0.000195  & 0.06197$\pm$0.0006 & 0.115111$\pm$0.0049  & 9.77$\pm$0.43     \\
   201128338  & K2-152      & 2 & 32.6479$\pm$0.01483  & 0.16952$\pm$0.00119  & 0.0344$\pm$0.0037  & 2.23$\pm$0.24   \\
   201132684.01  & K2-158b  & 2 & 5.90279$\pm$0.00233  & 0.06205$\pm$0.00068  & 0.0123$\pm$0.0012  & 1.27$\pm$0.13   \\
   201132684.02  & K2-158c  & 2 & 10.06049$\pm$0.00148  &  0.08853$\pm$0.00095 & 0.0255$\pm$0.0016  & 2.64$\pm$0.17   \\
   201155177  & K2-42       & 3  & 6.68796$\pm$0.00093  & 0.06339$\pm$0.00070  & 0.0304$\pm$0.0028  & 2.41$\pm$0.23   \\
   201160662  &             & 13 & 1.5374115$\pm$0.0000062  &  0.02800$\pm$0.00054    & 0.259$\pm$0.071     & 57.13$\pm$15.77  \\
   201264302  &             & 4 & 0.212194$\pm$0.000026  & 0.00532$\pm$0.00010  & 0.0271$\pm$0.004  & 1.25$\pm$0.18   \\ 
   201390927  &             & 2  & 2.638$\pm$0.0003  & 0.03585$\pm$0.00072  & 0.0265$\pm$0.0025  & 3.04$\pm$0.39   \\
   201393098  & K2-7        & 3 & 28.6777$\pm$0.0086  & 0.18752$\pm$0.00232  & 0.0177$\pm$0.0018  & 3.29$\pm$0.34   \\
   201403446  & K2-46       & 1  & 19.15454$\pm$0.002849  & 0.14283$\pm$0.00182  & 0.01705$\pm$0.00127  & 2.66$\pm$0.21   \\
   201407812  &             & 5 & 2.8268121            & 0.04192$\pm$0.00060  & 0.4560  & 119.51$\pm$4.02  \\
   201445732  &             & 13 & 11.20381$\pm$0.00055  &  0.09748$\pm$0.00122  &  0.0182$\pm$0.0027 & 2.37$\pm$0.35  \\
   201516974  &             &  6  & 36.7099$\pm$0.0125  & 0.23590$\pm$0.00833  & 0.0489$\pm$0.0033  & 31.18$\pm$2.50   \\
   201546283  & K2-27       & 1  & 6.771389$\pm$0.000062  & 0.06831$\pm$0.00071 & 0.049112$\pm$0.001573  & 4.70$\pm$0.16  \\
   201561956  &             & 13 & 13.2359$\pm$0.0031 &  0.10587$\pm$0.00162  &  0.0208$\pm$0.0046 & 2.17$\pm$0.49  \\
   201606542  &             & 4 & 0.444372$\pm$0.000042  & 0.01119$\pm$0.00011  & 0.0136$\pm$0.002  & 1.63$\pm$0.24   \\
   201649426  &             & 5   & 27.770388  & 0.16741$\pm$0.00090  & 0.3722  & 33.45$\pm$0.44  \\
   201754305.02  & K2-16b   & 3  & 7.61856$\pm$0.00096  & 0.06675$\pm$0.00071  & 0.0268$\pm$0.0022  & 1.93$\pm$0.16   \\
   201754305.01  & K2-16c   & 3  & 19.077$\pm$0.0033  & 0.12310$\pm$0.00131  & 0.0299$\pm$0.003  & 2.15$\pm$0.22    \\
   201779067  &             & 5 & 27.242912    & 0.19034$\pm$0.00326  & 0.2367  & 64.10$\pm$1.94 \\
   201841433  &             & 5  & 12.339133    & 0.09614$\pm$0.00097  & 0.02881  &  2.33$\pm$0.21 \\
   201855371  & K2-17       & 1 & 17.969079$\pm$0.0014  & 0.11508$\pm$0.00085  & 0.029715$\pm$0.003  & 1.96$\pm$0.20   \\
   201856786.01 &           & 13 & 3.83794$\pm$0.00041 & 0.04178$\pm$0.00090  &   0.0172$\pm$0.003 &  1.46$\pm$0.26 \\
   201856786.02 &           & 13 & 5.24086$\pm$0.00094  & 0.05143$\pm$0.00111 &   0.0166$\pm$0.0027 & 1.41$\pm$0.24 \\
   201912552  & K2-18       & 3 & 32.9418$\pm$0.0021  & 0.15444$\pm$0.01138  & 0.0517$\pm$0.0021  & 2.46$\pm$0.14  \\
   201923289  &             & 5 & 0.78214992    & 0.01616$\pm$0.00021  & 0.01346  & 1.34$\pm$0.25   \\
   202634963  &             & 5  & 28.707623     & 0.20176$\pm$0.00356  & 0.2136   &  44.32$\pm$1.25   \\
   202675839  &             & 1 & 15.466674$\pm$0.0016  & 0.13015$\pm$0.00205  & 0.12002$^{+0.3}_{-0.062}$  & 21.36$\pm$53.40  \\
   202821899  &             & 1  & 4.474513$\pm$0.0003  & 0.05944$\pm$0.00115  & 0.033719$\pm$0.0056  & 8.32$\pm$1.43   \\
   %   202900527  & K2-51    & 13.008383$\pm$0.00021  &   & 0.101010$\pm$0.00224  & 23.10$\pm$0.67 & \\ %FP Shporer+2017 removed from everywhere
   203070421  &             & 5  & 1.7359447     & 0.03340$\pm$0.00062  & 0.02551  & 7.66$\pm$0.81   \\
   203518244  &             & 5  & 0.8411257     & 0.01893$\pm$0.00019  & 0.01098  & 2.84$\pm$0.65  \\
   203533312  &             & 4 & 0.17566$\pm$0.000183  & 0.00698$\pm$0.00013  & 0.0248$\pm$0.001  &  7.23$\pm$0.35  \\
   203616858  &             & 13 & 1.68027$\pm$0.00011 & 0.02775$\pm$0.00051 &    0.0207$\pm$0.0238 & 2.85$\pm$3.28  \\
   203633064  &             & 13 & 0.7099504$\pm$0.0000013  & 0.01775$\pm$0.00020 &   0.357$\pm$0.079  &  82.26$\pm$18.41 \\
   203753577  &             & 5  & 3.4007758     & 0.04702$\pm$0.00077  & 0.06863  & 9.74$\pm$1.53  \\
   203771098.02  & K2-24b   & 1  & 20.885016$\pm$0.000438  & 0.15273$\pm$0.00097  & 0.045111$\pm$0.00227  & 5.71$\pm$0.30   \\
   203771098.01  & K2-24c   & 1  & 42.363982$\pm$0.000795  & 0.24473$\pm$0.00155  & 0.061091$\pm$0.00174  & 7.74$\pm$0.24  \\
   203826436.03  & K2-37b   & 1  & 4.443774$\pm$0.0005  & 0.05084$\pm$0.00056  & 0.017091$\pm$0.01883  & 1.56$\pm$1.72  \\
   203826436.01  & K2-37c   & 1  & 6.429582$\pm$0.0003  & 0.06503$\pm$0.00072  & 0.029105$\pm$0.00353  & 2.66$\pm$0.32   \\
   203826436.02  & K2-37d   & 1  & 14.090996$\pm$0.001078  & 0.10973$\pm$0.00121  & 0.027017$\pm$0.003572  &  2.47$\pm$0.33  \\
   203925865  &             & 13 & 8.796890$\pm$0.00059  & 0.08910$\pm$0.00084 &   0.0217$\pm$0.003 & 4.69$\pm$0.66  \\
   203929178  &             & 3  & 1.153886$\pm$0.000028  & 0.02385$\pm$0.00044  & 0.53$\pm$0.23  & 101.86$\pm$45.55   \\
   204197636  &             & 13 & 46.1373$\pm$0.00760 & 0.23732$\pm$0.00238  &    0.033$\pm$0.0024 & 3.06$\pm$0.2  \\
   204221263.02  & K2-38b   & 3  & 4.01628$\pm$0.00044  & 0.05009$\pm$0.00036  & 0.01329$\pm$0.00099  & 1.67$\pm$0.13   \\
   204221263.01  & K2-38c   & 3  & 10.56098$\pm$0.00081  & 0.09543$\pm$0.00068  & 0.0195$\pm$0.014  & 2.45$\pm$1.76   \\
   204914585  &             & 5 & 18.357773     & 0.14669$\pm$0.00221  & 0.01924  &  2.58$\pm$0.34 \\
   204991696  &             & 13 & 49.8558$\pm$0.0035 &  0.28089$\pm$0.00270  &     0.02222$\pm$0.0023 &  3.11$\pm$0.33 \\
   205071984.01  & K2-32b   & 1  & 8.991942$\pm$0.000158  & 0.08206$\pm$0.00084  & 0.056494$\pm$0.0013  &  5.19$\pm$0.14  \\
   205071984.03  & K2-32c   & 1  & 20.661623$\pm$0.001762  & 0.14289$\pm$0.00148  & 0.034033$\pm$0.001598  &  3.13$\pm$0.15   \\
   205071984.02  & K2-32d   & 1  & 31.715061$\pm$0.002567  & 0.19013$\pm$0.0019  & 0.037299$\pm$0.002528  & 3.43$\pm$0.24 \\
   205111664  &             & 5 & 15.937378    & 0.11803$\pm$0.00112  & 0.02135  & 2.24$\pm$0.27  \\
   205146011  &             & 13 & 1.057171$\pm$0.000061  & 0.01985$\pm$0.00023  &  0.0137$\pm$0.002 &  1.41$\pm$0.21 \\
   205170731  &             & 13 & 14.2005$\pm$0.0027  & 0.11034$\pm$0.00109   &    0.0276$\pm$0.0053 & 2.81$\pm$0.54  \\
   205470347  &             & 13 & 1.86732$\pm$0.00016  & 0.02727$\pm$0.00016  &   0.00857$\pm$0.00146  & 0.66$\pm$0.11  \\ % tiny AF
   205503762  &             & 13 & 6.4349$\pm$0.0012 &  0.06815$\pm$0.00085  &     0.0152$\pm$0.0052  &  2.24$\pm$0.77  \\
   205570849  &             & 3  & 16.8580$\pm$0.0011  & 0.12831$\pm$0.00168  & 0.047$\pm$0.057  & 6.21$\pm$7.53   \\
   205618538  &             & 13 & 2.167697$\pm$0.000022 & 0.03735$\pm$0.00081  &    0.04472$\pm$0.00154 & 11.38$\pm$0.57  \\
   205924614  & K2-55       & 3  & 2.849258$\pm$00.000033  & 0.03536$\pm$0.00031  & 0.0552$\pm$0.0013  & 4.17$\pm$0.11 \\
   205938820  &             & 13 & 4.20773$\pm$0.00075  & 0.04966$\pm$0.00052  &   0.0161$\pm$0.0023 & 1.53$\pm$0.22 \\
   205944181  &             & 1 & 2.475641$\pm$0.000057  & 0.03479$\pm$0.00042  & 0.055833$^{+0.19}_{-0.03}$  &  5.28$\pm$17.97  \\
   205950854  & K2-168      & 1 & 15.853989$\pm$0.001415  & 0.11803$\pm$0.00161  & 0.022489$\pm$0.001272  & 2.21$\pm$0.13   \\
   205951125  &             & 13 & 6.79143$\pm$0.0008 & 0.06487$\pm$0.00061   &     0.0259$\pm$0.0064  & 2.08$\pm$0.52  \\
   205957328  &             & 1 & 14.353438$\pm$0.001491  &  0.11117$\pm$0.00077  & 0.023912$\pm$0.004385  &  2.11$\pm$0.39  \\
   205998649  &             & 13 & 8.3958$\pm$0.0028 & 0.08268$\pm$0.00102  &     0.0181$\pm$0.007  & 3.87$\pm$1.50   \\
   206024342  &             & 3 & 14.637$\pm$0.0021   & 0.11259$\pm$0.00194  & 0.0249$\pm$0.0015  &  2.34$\pm$0.15  \\
   206026136  & K2-57       & 3  & 9.0063$\pm$0.0013  & 0.07525$\pm$0.00068  & 0.0308$\pm$0.0028  & 2.24$\pm$0.21  \\
   206036749  &             & 3 & 1.131316$\pm$0.00003  & 0.02226$\pm$0.00034  & 0.047$\pm$0.057  &  3.76$\pm$0.23  \\
   206038483  & K2-60       & 3 & 3.002627$\pm$0.000018  & 0.04178$\pm$0.00063  & 0.06191$\pm$0.00035   & 9.87$\pm$0.25  \\
   206047055  &             & 13 & 4.10290$\pm$0.00180 & 0.05208$\pm$0.00062  &    0.0106$\pm$0.0022 & 2.22$\pm$0.47  \\
   206055981  &             & 5 & 20.643928    & 0.12730$\pm$0.00099  & 0.03129  &  2.10$\pm$0.17  \\
   206082454.02  & K2-172b  & 1  & 14.316941$\pm$0.001445  & 0.11326$\pm$0.00110  & 0.017579$\pm$0.001495  & 1.67$\pm$0.14  \\
   206082454.01  & K2-172c  & 1 & 29.62682$\pm$0.001607  & 0.18392$\pm$0.00178  & 0.033824$\pm$0.001324  &  3.21$\pm$0.13  \\
   206103150.01  & WASP-47b & 3 & 4.159221$\pm$0.000015  & 0.05047$\pm$0.00058  & 0.10214$\pm$0.0003  & 12.71$\pm$0.27   \\
   206103150.02  & WASP-47d & 3 & 9.03164$\pm$0.00064  & 0.08464$\pm$0.00098  & 0.026$\pm$0.0015  &  3.24$\pm$0.20   \\
   206103150.03  & WASP-47e & 3 & 0.789518$\pm$0.00006  & 0.01667$\pm$0.00019  & 0.01344$\pm$0.00088  & 1.67$\pm$0.12  \\
   206114630  &             & 1 & 7.445026$\pm$0.0003  & 0.07031$\pm$0.00044  & 0.025337$\pm$0.033876  & 2.29$\pm$3.06   \\
   206125618  & K2-64       & 3 & 6.53044$\pm$0.00067  & 0.06671$\pm$0.00089  & 0.0259$\pm$0.0017  &   2.49$\pm$0.18  \\
   206135682  &             & 5 & 5.025831    & 0.05165$\pm$0.00037  & 0.01961  & 1.43$\pm$0.18 \\
   206208956  &             & 13 & 5.01038$\pm$0.00019  & 0.05878$\pm$0.00120  &    0.0257$\pm$0.0047  & 4.49$\pm$0.85  \\
   206245553  & K2-73       & 1  & 7.495692$\pm$0.000283   & 0.07520$\pm$0.00074  & 0.022901$\pm$0.001345  & 2.65$\pm$0.16   \\
   206260577  &             & 13 & 1.982116$\pm$0.000012 & 0.03254$\pm$0.00068  &   0.157$\pm$0.048 & 31.20$\pm$9.59  \\
   206369173  &             & 13 & 2.018725$\pm$0.000066 & 0.03656$\pm$0.00369  &   0.056$\pm$0.018  & 129.64$\pm$46.49  \\
   206414361  &             & 13 &  3.47722$\pm$0.00038 & 0.03675$\pm$0.00023  &    0.0253$\pm$0.0086 & 1.44$\pm$0.49  \\
   206417197  &             & 4   & 0.442094$\pm$0.000086  & 0.01071$\pm$0.00011  & 0.0138$\pm$0.001  & 1.18$\pm$0.09   \\ 
%   206432863  &          & 23.9726$\pm$0.0011        &    & 0.0793$\pm$0.0017   & 16.63$\pm$0.71   &   \\ %FP Shporer+2017 removed from everywhere
   206476150  &             & 13 & 12.19649$\pm$0.00082 & 0.10263$\pm$0.00120  &    0.0192$\pm$0.0019 & 2.10$\pm$0.21  \\
   210394706.02 &           & 13 & 3.16363$\pm$0.00029  & 0.03565$\pm$0.00025  &   0.0222$\pm$0.00380 & 1.41$\pm$0.24  \\
   210394706.01 &           & 13 & 15.0818$\pm$0.0025  & 0.10097$\pm$0.00070   &    0.0326$\pm$0.0045 & 2.08$\pm$0.29 \\
   210402237  & K2-79       & 1 & 10.993948$\pm$0.000627  & 0.09707$\pm$0.00101  & 0.027782$\pm$0.001543  & 3.85$\pm$0.22  \\
   210414957  &             & 3 & 0.969967$\pm$0.000012  & 0.02049$\pm$0.00020  &  0.35$\pm$0.15   & 80.64$\pm$34.62 \\
   210508766.01  & K2-83b   & 3  & 2.74697$\pm$0.00018  & 0.03182$\pm$0.00018  & 0.0268$\pm$0.0019  & 1.59$\pm$0.11  \\
   210508766.02  & K2-83c   & 3  & 9.99767$\pm$0.00081  & 0.07530$\pm$0.00043  & 0.0319$\pm$0.0018  & 1.89$\pm$0.11  \\
   210559259    &           & 7 &    14.2683$\pm$0.0012   & 0.10583$\pm$0.00105     & 0.02854$^{+0.0011}_{-0.00082}$ &  2.24$\pm$0.09 \\
   210609658  &             & 1   & 14.145239$\pm$0.000468  & 0.12894$\pm$0.00310  & 0.06327$\pm$0.00188  & 22.66$\pm$0.91   \\
   210629082  &             & 1 & 27.353103$\pm$0.007472  & 0.19187$\pm$0.00358  & 0.019308$\pm$0.0029   & 4.13$\pm$0.63  \\
   210664763  &             & 13 & 3.72007$\pm$0.00047  & 0.04714$\pm$0.00064  &   0.01450$\pm$0.003 &  1.56$\pm$0.32 \\
   210678858.03  &          & 13 & 10.0696$\pm$0.0013 & 0.08767$\pm$0.00066  &     0.0190$\pm$0.0033  & 1.66$\pm$0.29  \\
   210678858.02  &          & 13 & 14.8484$\pm$0.0011 & 0.11358$\pm$0.00085  &     0.0302$\pm$0.003  & 2.64$\pm$0.26  \\
   210678858.01  &          & 13 & 31.3537$\pm$0.0019 & 0.18695$\pm$0.00140  &    0.0432$\pm$0.003  & 3.78$\pm$0.27  \\
   210707130  & K2-85       & 1 & 0.684553$\pm$0.000013  & 0.01348$\pm$0.00011  & 0.018081$\pm$0.001436  &  1.32$\pm$0.11  \\
   210718708  & K2-86       & 1  & 8.775864$\pm$0.0009   & 0.07978$\pm$0.00093  & 0.025082$\pm$0.003131  & 2.27$\pm$0.28  \\
   210731500  & K2-87       & 3  & 9.72739$\pm$0.00087  & 0.08914$\pm$0.00124  & 0.0441$\pm$0.0032  & 6.79$\pm$0.51  \\
   210775710  &             & 1 & 59.848566$\pm$0.000184  & 0.29810$\pm$0.00401  & 0.100817$\pm$0.001863  & 11.45$\pm$0.27  \\
   210857328  & K2-177      & 1 & 14.155185$\pm$0.00315  & 0.12655$\pm$0.00223  & 0.015987$\pm$0.0018  & 3.07$\pm$0.36  \\
   210961508  &             & 4 & 0.349935$\pm$0.000042  &  0.01050$\pm$0.00036  & 0.0263$\pm$0.003   &  8.47$\pm$1.01  \\
   211087003.02  &          & 13 & 28.29213$\pm$0.00126 & 0.18102$\pm$0.00229  &    0.0338$\pm$0.0023 & 3.84$\pm$0.27  \\
   211327855  &             & 13 & 1.72397$\pm$0.00027  & 0.02727$\pm$0.00028  &   0.0137$\pm$0.0038 & 1.26$\pm$0.35  \\
   211335816  &             & 8   & 4.99    & 0.06106$\pm$0.00103   & 0.043667$\pm$0.0025   & 8.25$\pm$0.53   \\ 
   211336616  &             & 8 & 44.13   & 0.26941$\pm$0.02413   & 0.020655$\pm$0.0025   & 25.26$\pm$3.81   \\ 
   211351816  & K2-97       & 1  & 8.405276$\pm$0.001166  &  0.09382$\pm$0.00307  & 0.025002$\pm$0.003158  & 12.07$\pm$1.66  \\
   211355342  & K2-181      & 1 & 6.894252$\pm$0.00043   & 0.07088$\pm$0.00085  & 0.024829$\pm$0.002084  & 2.87$\pm$0.25   \\
   211357309  &             & 9 & 0.46395$\pm$0.00002  & 0.00921$\pm$0.00005  & 0.017$\pm$0.001  &  0.86$\pm$0.05     \\ % tiny AF
   211359660  & K2-182      & 1 & 4.736884$\pm$0.000075  & 0.05257$\pm$0.00046  & 0.032108$\pm$0.001498  & 2.77$\pm$0.13 \\
   211365543  &             & 8 & 5.264    & 0.06275$\pm$0.00082  & 0.009804    &  1.68$\pm$0.43  \\
   211390903  &             & 10  & 7.757595$\pm$0.000822  & 0.09205$\pm$0.00502  & 0.0251$\pm$0.0007  & 30.42$\pm$1.76   \\ 
   211491383  & K2-269 & 1  & 4.145398$\pm$0.001032  & 0.05213$\pm$0.00100  & 0.008372$\pm$0.001162  & 1.34$\pm$0.20  \\
   211535327  &             & 13 & 20.2244$\pm$0.0021  & 0.13749$\pm$0.00166  &    0.0323$\pm$0.0043 & 3.03$\pm$0.41 \\
   211562654.03  & K2-183b  & 1  & 0.469269$\pm$0.000026  & 0.01139$\pm$0.00014  & 0.027288$^{+0.27}_{-0.015}$  & 2.88$\pm$28.54 \\
   211562654.01  & K2-183c  & 1 & 10.793471$\pm$0.000803  & 0.09213$\pm$0.00117  & 0.026365$\pm$0.002542  & 2.79$\pm$0.27    \\
   211562654.02  & K2-183d  & 1 & 22.629496$\pm$0.001949  & 0.15093$\pm$0.00192  & 0.026677$\pm$0.002712  & 2.82$\pm$0.29   \\
   211586387  &             & 8 & 35.383  & 0.22064$\pm$0.00402  & 0.18841$\pm$0.00165  & 2.25$\pm$0.19    \\
   211611158.02  &          & 1   & 52.714072$\pm$0.003819  & 0.27437$\pm$0.00257  & 0.02803$\pm$0.00436  & 2.79$\pm$0.44     \\
   211611158  & K2-185b     & 1 & 10.616646$\pm$0.0018  & 0.09427$\pm$0.00089  & 0.013164$\pm$0.002118  & 1.31$\pm$0.21    \\
   211733267  &             & 1   & 8.658168$\pm$0.00003  & 0.07925$\pm$0.00083  & 0.1921$^{+0.114}_{-0.059}$  & 18.94$\pm$11.25    \\
   211736305  &             & 13 & 14.5616$\pm$0.0026 & 0.11075$\pm$0.00138  &     0.0305$\pm$0.0149 & 2.71$\pm$1.33 \\
   211736671  & K2-108      & 1 & 4.73379$\pm$0.000153  & 0.05695$\pm$0.00065  & 0.030069$\pm$0.002987  & 5.75$\pm$0.59    \\
   211763214  &             & 1 & 21.191788$\pm$0.003275  & 0.14294$\pm$0.00129  & 0.015441$\pm$0.00162  & 1.35$\pm$0.14    \\
   211770696  &             & 1  & 16.27284$\pm$0.002441  & 0.12608$\pm$0.00175  & 0.018155$\pm$0.00156  &  2.66$\pm$0.24    \\
   211800191  &             & 1  & 1.106175$\pm$0.000009  & 0.02092$\pm$0.00040  & 0.089351$\pm$0.06  & 11.42$\pm$7.67    \\
   211816003  & K2-272      & 11 & 14.453513$\pm$0.001783  & 0.10872$\pm$0.00145  & 0.0336$\pm$0.0041  & 2.98$\pm$0.37    \\
   211818569  & K2-121      & 1  & 5.185759$\pm$0.000014  & 0.05269$\pm$0.00037  & 0.10208$\pm$0.003964  & 7.49$\pm$0.30    \\
   211923431  &             & 8 & 29.729      & 0.18570$\pm$0.00199  & 0.025878$\pm$0.0025  & 3.28$\pm$0.33    \\
   211945201  &             & 1 & 19.491795$\pm$0.000516  & 0.14891$\pm$0.00228  & 0.038014$\pm$0.002554  &  5.81$\pm$0.40    \\
   211970147  & K2-102      & 12 & 9.915651$\pm$0.001194  & 0.08342$\pm$0.00073  & 0.0169$\pm$0.001  & 1.35$\pm$0.08  \\
   211978988  &             & 1  & 36.556251$\pm$0.004239  &  0.21767$\pm$0.00283  & 0.026283$\pm$0.001964  & 3.24$\pm$0.25   \\
   211990866  & K2-100      & 12 & 1.673915$\pm$0.000011  & 0.02882$\pm$0.00028  & 0.0267$\pm$0.0011  & 3.64$\pm$0.16    \\
   212006344  & K2-122      & 9 & 2.21940$\pm$0.00007  & 0.02828$\pm$0.00020  & 0.020$\pm$0.001  & 1.29$\pm$0.07    \\
   212099230  &             & 11 & 7.112273$\pm$0.000284  & 0.07139$\pm$0.00131  & 0.0302$\pm$0.0011  & 3.19$\pm$0.12    \\
   212110888  & K2-34       & 1  & 2.995646$\pm$0.000006  & 0.04285$\pm$0.00076  & 0.088002$\pm$0.001666  & 13.93$\pm$0.39   \\
   212136123  &             & 8 & 2.226     & 0.03192$\pm$0.00033  & 0.026003$\pm$0.0025  & 2.27$\pm$0.22   \\
   212141021  &             & 8  & 2.918     & 0.03729$\pm$0.00041  & 0.015674$\pm$0.0025  & 1.33$\pm$0.21   \\
   212159623  &             & 13 & 4.70751$\pm$0.00065  & 0.05533$\pm$0.00078   &   0.0139$\pm$0.002 & 1.51$\pm$0.22 \\
   212164470.01  & K2-188b  & 1 & 1.742983$\pm$0.00026   & 0.02881$\pm$0.00041  & 0.010407$\pm$0.0009  & 1.36$\pm$0.12   \\
   212164470.02  & K2-188c  & 1 & 7.807595$\pm$0.000597  & 0.07827$\pm$0.00112  & 0.021697$\pm$0.001430  & 2.84$\pm$0.20   \\
   212300977  & WASP-55     & 11 & 4.465635$\pm$0.000023  & 0.05359$\pm$0.00058  & 0.1223$\pm$0.0004  & 15.09$\pm$0.26   \\
   212301649  &             & 8 & 1.225      & 0.02145$\pm$0.00031  & 0.014962$\pm$0.0025  & 1.40$\pm$0.25   \\
   212362217  &             & 13 & 0.6962935$\pm$0.0000087 & 0.01514$\pm$0.00027  &  0.0319$\pm$0.0369 &  3.94$\pm$4.56  \\
   212393193.01  &          & 8   & 14.452  & 0.11948$\pm$0.00141  & 0.0182$\pm$0.0025  & 2.29$\pm$0.32    \\ 
   212393193.02  &          & 8 & 36.152  & 0.22018$\pm$0.00259  & 0.0183$\pm$0.0025  & 2.30$\pm$0.32    \\
   212425103  &             & 8 & 0.946      & 0.01782$\pm$0.00024  & 0.017346$\pm$0.0025  &  1.54$\pm$0.23    \\
   212432685  &             & 11  & 0.531704$\pm$0.000035  & 0.01293$\pm$0.00021  & 0.0169$\pm$0.0018  & 2.18$\pm$0.43   \\
   212440430  &             & 8 & 19.991     & 0.14224$\pm$0.00187  & 0.023276$\pm$0.0025  & 2.54$\pm$0.28    \\
   212464382  &             & 13 & 4.07337$\pm$0.00051 & 0.04757$\pm$0.00046  &   0.01071$\pm$0.00184 & 0.94$\pm$0.16  \\
   212495601  &             & 8  & 21.677     & 0.14710$\pm$0.00177  & 0.024596$\pm$0.0025  & 2.71$\pm$0.28   \\
   212521166  & K2-110      & 1 & 13.863910$\pm$0.000229  & 0.10373$\pm$0.00085  & 0.033432$\pm$0.001766  & 2.61$\pm$0.14   \\
   212560683  &             & 13 & 13.7043$\pm$0.0037 & 0.11317$\pm$0.00114   &    0.0118$\pm$0.0033 & 1.31$\pm$0.37   \\ 
   212585579  &             & 11 & 3.021795$\pm$0.000094  &  0.04170$\pm$0.00056  & 0.3876$\pm$0.3569  & 46.56$\pm$42.88    \\
   212587672  &             & 1 & 23.226001$\pm$0.003092  & 0.15929$\pm$0.00198  & 0.021599$\pm$0.003624  & 2.33$\pm$0.39   \\
   212624936  &             & 13 & 11.81387$\pm$0.00093 & 0.09971$\pm$0.00128  &    0.0258$\pm$0.0036 &  2.63$\pm$0.37 \\
   212639319  &             & 1  & 13.843725$\pm$0.000948  &  0.12740$\pm$0.00167  & 0.037754$^{+0.297}_{-0.0096}$  & 11.05$\pm$86.92   \\  % WTF error bar
   212645891  &             & 1 & 0.328152$\pm$0.000001  & 0.00934$\pm$0.00018  & 0.136972$^{+0.113}_{-0.06}$  & 17.05$\pm$14.07   \\
   212646483  &             & 8  & 8.253      & 0.08348$\pm$0.00122  & 0.029071$\pm$0.0025  & 6.98$\pm$0.66   \\
   212652418  &             & 13 & 19.1324$\pm$0.0031 & 0.14091$\pm$0.00202  &    0.0186$\pm$0.0022 & 2.78$\pm$0.34  \\
   212672300  & K2-194      & 1 & 39.721386$\pm$0.0057  & 0.24073$\pm$0.00258  & 0.026065$\pm$0.002509  & 3.90$\pm$0.39   \\
   212686205  & K2-128      & 1 & 5.675814$\pm$0.000427  & 0.05520$\pm$0.00050  & 0.016952$\pm$0.00133  & 1.22$\pm$0.10   \\
   212688920  &             & 8 & 62.841  & 0.30670$\pm$0.00604  & 0.231222$\pm$0.0025  & 27.02$\pm$0.62   \\
   212689874.01  & K2-195b  & 1  & 15.853543$\pm$0.00079  & 0.12127$\pm$0.00172  & 0.029741$\pm$0.001265  &  3.20$\pm$0.15   \\
   212689874.02  & K2-195c  & 1 & 28.482786$\pm$0.00731  & 0.17922$\pm$0.00257  & 0.026054$\pm$0.0024  & 2.81$\pm$0.26   \\
   212779596.01  & K2-199b  & 1  & 3.225423$\pm$0.000071  & 0.03811$\pm$0.00035  & 0.025852$\pm$0.002447  & 1.89$\pm$0.18   \\
   212779596.02  & K2-199c  & 1 & 7.374497$\pm$0.000118  & 0.06614$\pm$0.00060  & 0.038968$\pm$0.002060  & 2.86$\pm$0.15  \\
   212803289  & K2-99       & 1  & 18.248708$\pm$0.000634  & 0.15352$\pm$0.00168  & 0.042431$\pm$0.001169  & 12.42$\pm$0.48   \\
   212828909  & K2-200      & 1  & 2.849883$\pm$0.000188  & 0.03724$\pm$0.00027  & 0.015799$\pm$0.001590  & 1.33$\pm$0.13   \\
   213408445  &             & 13 & 2.49686$\pm$0.00022  & 0.04315$\pm$0.00386  &  0.072$\pm$0.022 &  301.12$\pm$94.83 \\
   213546283  &             & 1 & 9.770186$\pm$0.000325  & 0.08877$\pm$0.00103  & 0.029436$\pm$0.0015  & 3.73$\pm$0.20  \\
   213703832  &             & 11   & 0.515513$\pm$0.000024  & 0.01397$\pm$0.00157  & 0.0409$\pm$0.0096  & 50.02$\pm$13.08  \\
   213840781  &             & 11 & 12.364531$\pm$0.000375  & 0.10365$\pm$0.00208  & 0.4363$\pm$0.2602   & 60.98$\pm$36.40  \\
   214419545  &             & 13 & 9.40172$\pm$0.00048 & 0.08572$\pm$0.00096  &   0.016$\pm$0.0021 & 2.36$\pm$0.31  \\
   214630761  &             & 13 & 1.236438$\pm$0.000022 & 0.02620$\pm$0.00050  &  0.143$\pm$0.04 &  48.41$\pm$13.94 \\
   214741009  &             & 11 & 7.269622$\pm$0.000521  & 0.09463$\pm$0.00400  & 0.4156$\pm$0.3808   & 419.79$\pm$386.31   \\ % LOL
   214888033  &             & 13 & 7.457597$\pm$0.000096 & 0.07353$\pm$0.00086  &  0.077$\pm$0.015 &  9.42$\pm$1.84 \\
   214984368  &             & 13 & 0.2633809$\pm$0.000003 & 0.01066$\pm$0.00119  &  0.090$\pm$0.021 & 440.29$\pm$137.57  \\ 
   215125108  &             & 13 & 0.738067$\pm$0.000026 & 0.01837$\pm$0.00149  &  0.095$\pm$0.027 &  232.37$\pm$73.82 \\
   215175768  &             & 13 & 1.726115$\pm$0.000098 & 0.02788$\pm$0.00040  &  0.0610$\pm$0.021 &  6.28$\pm$2.17 \\
   215364084  &             & 13 & 2.74324$\pm$0.00017  & 0.04290$\pm$0.00192  &  0.0526$\pm$0.0265 &  30.33$\pm$15.39 \\
   215381481  &             & 13 & 0.533393$\pm$0.000027 & 0.01352$\pm$0.00096   &  0.01206$\pm$0.00232 & 72.96$\pm$15.99  \\
   216111905  &             & 13 & 3.02030$\pm$0.00032 & 0.04040$\pm$0.00045   &   0.0410$\pm$0.020 & 5.73$\pm$2.80  \\
   216363472  &             & 13 & 8.69290$\pm$0.00085  & 0.08138$\pm$0.00108  &  0.0154$\pm$0.0170 & 1.68$\pm$1.86  \\
   216405287  & K2-202      & 1 & 3.405164$\pm$0.000126   & 0.04334$\pm$0.00061  & 0.023171$\pm$0.001335  & 2.28$\pm$0.14    \\
   216494238  & K2-280      & 1 & 19.894641$\pm$0.002898  & 0.14649$\pm$0.00183  & 0.047857$\pm$0.002267  & 6.74$\pm$0.35    \\
   218195416  &             & 13 & 0.4951253$\pm$0.0000031 & 0.01447$\pm$0.00023  & 0.1410$\pm$0.0130 & 33.41$\pm$3.54  \\
   218300572  &             & 13 & 1.589843$\pm$0.000013 & 0.03266$\pm$0.00094  &  0.114$\pm$0.033 & 43.20$\pm$12.86  \\
   219388192  &             & 1  & 5.292605$\pm$0.000031  & 0.05860$\pm$0.00076  & 0.094335$\pm$0.000852  &  10.92$\pm$0.22   \\
   219480273  &             & 13 & 26.48370$\pm$0.0051 & 0.17671$\pm$0.00195   &    0.0132$\pm$0.0033 &  2.03$\pm$0.51 \\
   219800881  &  K2-231     & 13 & 13.84457$\pm$0.00154 & 0.11357$\pm$0.00156  &   0.0248$\pm$0.0018 & 2.74$\pm$0.20  \\
   220170303  & K2-203      & 1 & 9.695101$\pm$0.001334  & 0.08375$\pm$0.00062  & 0.01647$\pm$0.003246  &  1.37$\pm$0.27   \\
   220186645  & K2-204      & 1 & 7.055784$\pm$0.000650  & 0.07246$\pm$0.00090  & 0.023711$\pm$0.00094  & 3.50$\pm$0.18     \\
   220198551  &             & 13 & 0.7988453$\pm$0.0000083 & 0.01593$\pm$0.00011   & 0.079$\pm$0.035 &  7.26$\pm$3.22 \\
   220209578  &             & 11 & 8.904519$\pm$0.000205  & 0.08322$\pm$0.00115  & 0.3805$\pm$0.3287  &  44.04$\pm$38.07  \\
   220245303  &             & 1 & 3.680340$\pm$0.000359  & 0.04394$\pm$0.00032  & 0.012565$\pm$0.0022  &  1.05$\pm$0.18    \\ % Earthy McEarthface?
   220282718  &             & 13 & 0.5551606$\pm$0.0000058 & 0.01364$\pm$0.00019  & 0.0630$\pm$0.034 & 11.21$\pm$6.06  \\
   220322327  &             & 13 & 3.313470$\pm$0.00024  & 0.04074$\pm$0.00047   &  0.042$\pm$0.033 &  3.62$\pm$2.85 \\
   220341183  & K2-213      & 1 & 8.130870$\pm$0.001799  & 0.08241$\pm$0.00088  & 0.011526$\pm$0.001564  & 1.66$\pm$0.23   \\
   220400100 &              & 7  &    10.7946$\pm$0.0019   & 0.08817$\pm$0.00080  & 0.0314$^{+0.0039}_{-0.0019}$ & 2.49$\pm$0.31  \\ 
   220431824  &             & 13 & 9.073266$\pm$0.000037 & 0.08652$\pm$0.00102  &  0.1213$\pm$0.0026 & 23.71$\pm$0.71  \\
   220436189  &             & 13 & 13.60940$\pm$0.00330 &  0.09313$\pm$0.00064    &   0.0396$\pm$0.0045 &   2.42$\pm$0.28 \\
   220436208  &             & 11  & 5.235714$\pm$0.000316  & 0.05920$\pm$0.00071  & 0.0337$\pm$0.0034  &  4.38$\pm$0.46   \\
   220459477  &             & 13 & 2.38098$\pm$0.00018  & 0.03250$\pm$0.00039  &  0.0215$\pm$0.0038 & 1.82$\pm$0.32  \\
   220470563  &             & 13 & 7.30383$\pm$0.00043 & 0.06855$\pm$0.00049   &   0.02790$\pm$0.0036 & 2.23$\pm$0.29  \\
   220481411  & K2-216      & 1 & 2.174789$\pm$0.000039   & 0.02953$\pm$0.00029  & 0.023117$\pm$0.001166  & 1.74$\pm$0.09    \\
   220621788  & K2-220      & 1 & 13.682511$\pm$0.000721  & 0.10864$\pm$0.00103  & 0.021843$\pm$0.001610  & 2.43$\pm$0.18   \\
   220629489  & K2-283      & 11 & 1.921076$\pm$0.000050  & 0.02890$\pm$0.00028  & 0.0404$\pm$0.0048  &  3.59$\pm$0.43    \\
   220639177  &             & 13 & 7.14238$\pm$0.00069  & 0.06660$\pm$0.00067   &  0.0236$\pm$0.0057 &  1.86$\pm$0.45  \\
   220643470  &             & 1 & 2.653230$\pm$0.000089  & 0.04349$\pm$0.00461  & 0.041582$\pm$0.002685  & 134.86$\pm$21.21   \\
   220674823.01  &          & 1  & 0.571299$\pm$0.000015  & 0.01329$\pm$0.00017  & 0.016876$\pm$0.00137  & 1.82$\pm$0.15     \\
   220674823.02  &          & 1 & 13.339746$\pm$0.001089  & 0.10854$\pm$0.00138  & 0.027358$\pm$0.003262  & 2.95$\pm$0.35   \\
   228725791.01  & K2-247b  & 2 & 2.25021$\pm$0.00036  & 0.02989$\pm$0.00027  & 0.0283$\pm$0.0025  & 2.10$\pm$0.19   \\
   228725791.02  & K2-247c  & 2  & 6.49424$\pm$0.00260  & 0.06059$\pm$0.00056  & 0.0292$\pm$0.0032  &  2.17$\pm$0.24  \\
   228734889  &             & 1 & 48.249552$\pm$0.000173  & 0.25637$\pm$0.00368  & 0.172572$\pm$0.00245  & 19.60$\pm$0.53   \\
   228735255  & K2-140      & 1 & 6.569213$\pm$0.000020  & 0.06909$\pm$0.00089  & 0.114173$\pm$0.000560  & 12.72$\pm$0.32  \\
   228736155  & K2-226      & 1 & 3.271106$\pm$0.000369  & 0.04227$\pm$0.00071  & 0.016535$\pm$0.001862  & 1.66$\pm$0.19   \\
   228754001  & K2-132      & 1 & 9.173866$\pm$0.001534  & 0.09237$\pm$0.00294  & 0.029103$\pm$0.001475  & 12.43$\pm$0.73   \\
   229017395  & K2-258      & 2 & 19.09210$\pm$0.00633  & 0.13931$\pm$0.00174  & 0.0210$\pm$0.0014  &  3.12$\pm$0.22  \\
   247047370 &              & 7 & 4.20566$\pm$0.00018 & 0.04910$\pm$0.00064    & 0.0267$\pm$0.0029 & 2.50$\pm$0.27  \\ 
   247063356 &              & 7  & 9.7051$\pm$0.0016 & 0.09163$\pm$0.00119   & 0.0197$\pm$0.0020 & 2.37$\pm$0.24  \\ 
   \hline
    \end{tabular}
    \caption{Planet-candidate properties.  References -- 1: \citet{mayo18}, 2: \citet{l18b}, 3: \citet{c16}, 4: \citet{adams16}, 5: \citet{v16}, 6: \citet{schmitt16}, 7: \citet{zink19}, 8: \citet{pope16}, 9: \citet{dressing17}, 10: \citet{nardiello16}, 11: \citet{petigura17}, 12: \citet{mann17}, 13: \citet{kruse19} }
    \label{tab:planetparams}
\end{table*} 

Using our self-consistent stellar radii, we find the derived planet-candidate radii to lie in a reasonable range for approximately 90\% of the planet candidates examined here.  We set an upper limit of 2\Rjup\ (22\,\Rearth), a radius larger than which no planet has been confirmed.  By this criterion, we find 30 candidates with unphysically large radii, and we strongly suspect them to be false positives.  All have a disposition status of "candidate" (i.e. not "confirmed") on the NASA Exoplanet Archive, and they are enumerated in Table~\ref{tab:heftybois}.  

\begin{table*}
    \centering
    \begin{tabular}{lrl}
    \hline
  EPIC   &    $R_p$ (\Rearth) & Comments \\ 
    \hline
201160662   &  57.13$\pm$15.77   &  \textit{Gaia} RV error 4.5$\sigma$ too large  \\
201407812   &   119.51$\pm$4.02  & Double-lined binary. \textit{Gaia} RV error 3.0$\sigma$ too large \\
201516974   &   31.18$\pm$2.50 & \textit{Gaia} RV error 4.0$\sigma$ too large. Seismic log $g$=2.934$\pm$0.010 \\
201649426   &   33.45$\pm$0.44 & \textit{Gaia} RV error 4.6$\sigma$ too large \\
201779067   &     64.10$\pm$1.94 & \textit{Gaia} RV error 8.2$\sigma$ too large \\
202634963   &      44.32$\pm$1.25 & Double-lined binary \\
203633064   &   82.26$\pm$18.41  &      \\  % inexplicable shit Kruse candidate
203929178   &        101.86$\pm$74.42 & \textit{Gaia} astrometric noise 419$\sigma$  \\
206260577   &      31.20$\pm$9.59 &    \\  
206369173   &     129.64$\pm$46.49 &  log $g$=1.69$\pm$0.15  \\   
210414957   &       80.64$\pm$34.62  & Large uncertainty from $R_p/R_*$ \\
210609658   &      22.66$\pm$0.91 & \textit{Gaia} RV error 3.1$\sigma$ too large \\
211336616   &      25.26$\pm$3.81  & log $g$=2.06$\pm$0.18  \\ 
211390903   &     30.42$\pm$1.76  & log $g$=2.89$\pm$0.19. Seismic log $g$=2.626$\pm$0.022 \\
212585579   &   46.56$\pm$42.88 & \textit{Gaia} RV error 3.1$\sigma$ too large  \\
212688920   &    27.02$\pm$0.62  &      \\   % not even found in Gaia???
213408445   &     301.12$\pm$94.83 & log $g$=1.21$\pm$0.19    \\ 
213703832   &       50.02$\pm$13.08 & log $g$=2.34$\pm$0.21. \\
213840781   &        60.98$\pm$36.40  & Large uncertainty from $R_p/R_*$. \\
214630761   &      48.41$\pm$13.94 &    \\ % not even found in Gaia???
214741009   &   419.79$\pm$386.31  & log $g$=2.25$\pm$0.21. \\
214984368   &  440.29$\pm$137.57  & log $g$=1.50$\pm$0.18  \\ 
215125108   &  232.37$\pm$73.82   &  log $g$=2.01$\pm$0.21  \\  
215364084   &    30.33$\pm$15.39   &  log $g$=3.08$\pm$0.22 \\  
215381481   &    72.96$\pm$15.99  &  log $g$=0.73$\pm$0.19 \\  
218195416   &    33.41$\pm$3.54   &   \\   
218300572   &    43.20$\pm$12.86  &    \\  % not even found in Gaia???
220209578   &   122.05$\pm$105.52  & Large uncertainty from $R_p/R_*$.  \\
220431824   &    23.71$\pm$0.71  &  \\  
220643470   &   134.86$\pm$21.21  & log $g$=1.51$\pm$0.13 \\  
   \hline
    \end{tabular}
    \caption{Candidates larger than 22\,\Rearth.  These candidates are highly likely to be false positives. }
    \label{tab:heftybois}
\end{table*} 

We checked the \textit{Gaia} DR2 results for evidence of hidden binarity in these 30 targets.  One star (EPIC 203929178) had highly significant excess astrometric noise (hundreds of sigma).  A further seven stars had uncertainties in their absolute radial velocities more then $3\sigma$ larger than the expected RV precision for stars of their temperature \citep{katz19}.  We also flag eleven stars as giants with log\,$g \ltsimeq 3.0$ from our spectroscopic determination.  Those giant-star hosts are more likely to be false positives, e.g. wherein a grazing eclipse by an M dwarf can produce the $K2$ transit-like signal, or where the transiting object orbits a different star, as postulated by the analysis of \textit{Kepler} giants in \citet{sliski14}.  Two stars have a weak secondary set of spectral lines, and are marked as binaries here.  None of the 30 stars in Table~\ref{tab:heftybois} have K2-HERMES-derived stellar parameters that are unusually imprecise (Table~\ref{tab:stellarparams}), and so we are confident in our disposition of these planetary candidates as false positives due to their unrealistically large inferred radii.  Furthermore, two stars in Table~\ref{tab:heftybois} have seismic detections confirming their evolved nature.  EPIC\,211351816, hosting the confirmed planet K2-97b \citep{grunblatt18}, also has a seismic detection.  We derive its radius to be 4.11$\pm$0.07\,\Rsun\ (Table~\ref{tab:seismology}), in turn yielding a planetary radius of 11.22$\pm$1.43\,\Rearth\ which agrees with our K2-HERMES radius determination (12.07$\pm$1.66\,\Rearth), and is about 3$\sigma$ smaller than the radius given by \citet{grunblatt18}. 

% we get 12.07 +/- 1.66 R_e
% Grunblatt+18 K2-97 R = 14.7 +/- 1.2 R_e, mass = 0.48 +/- 0.07 Mjup.
%[AGAIN, FOR THOSE OF THEM THAT ARE GIANTS, WE CAN DOUBLE CHECK WITH SEISMOLOGY].  

\begin{figure*}
\includegraphics[width=\columnwidth]{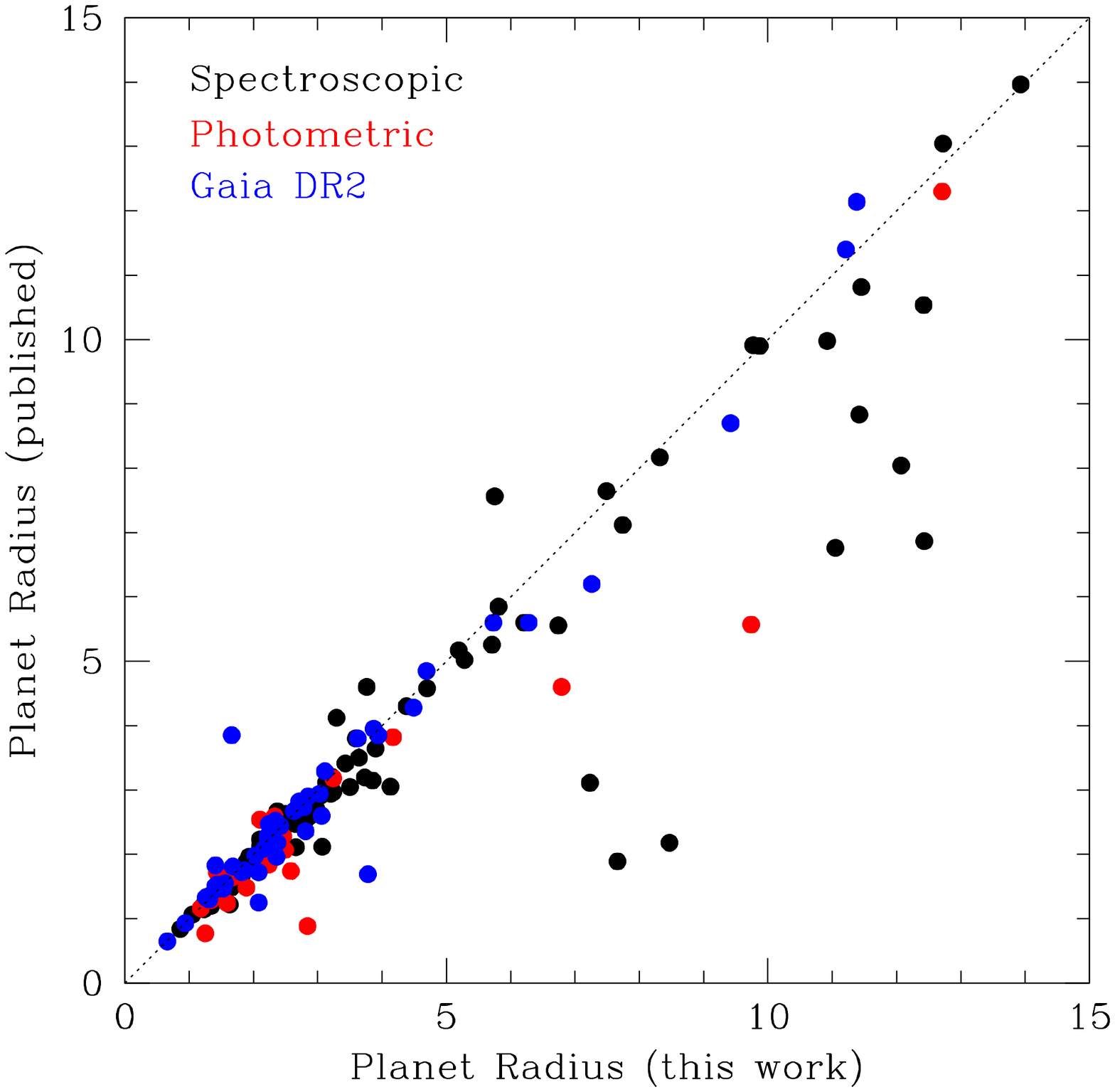}
\includegraphics[width=\columnwidth]{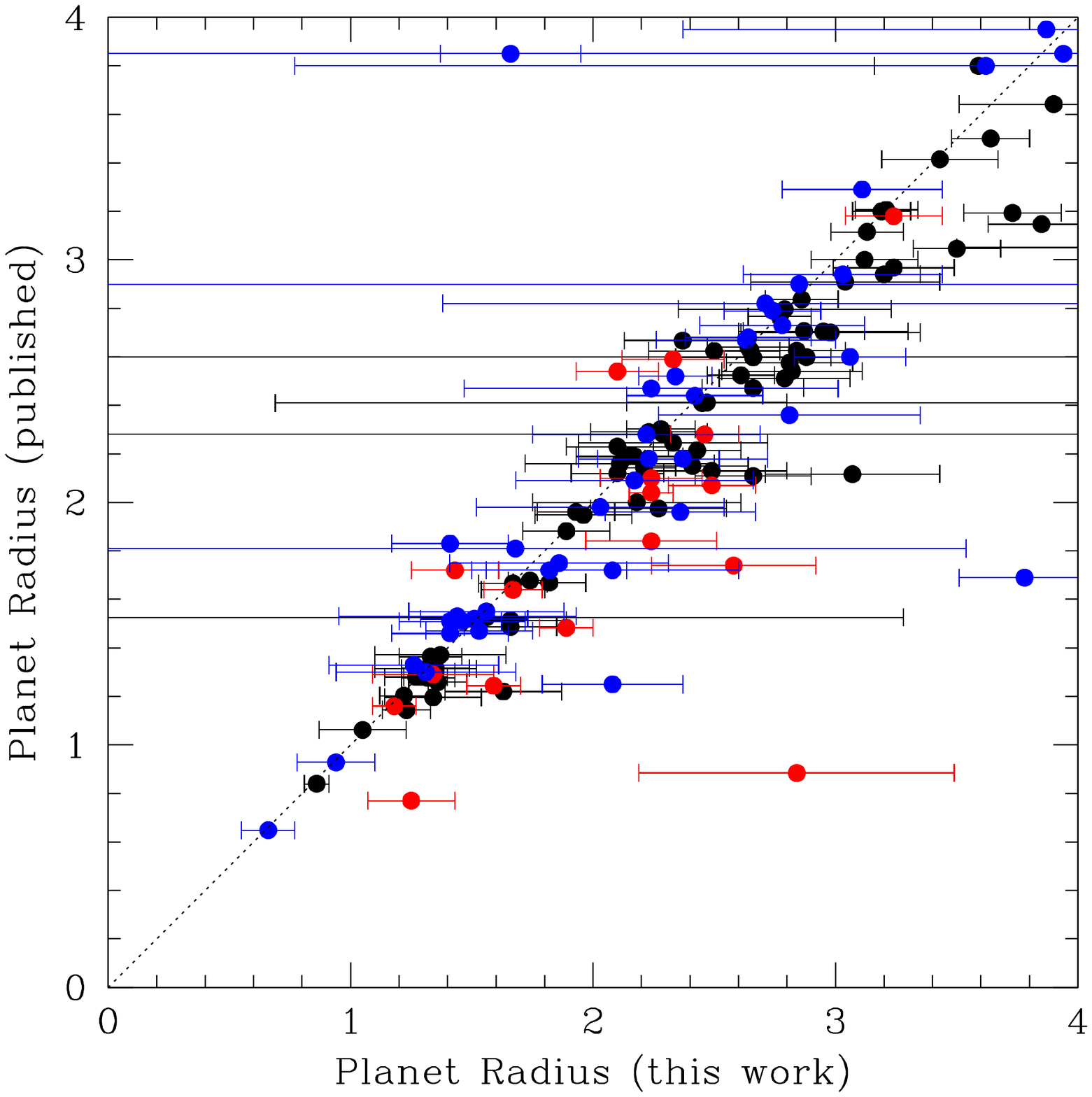}
\caption{Left panel: Comparison of our derived planetary radii with those from the literature.  Error bars have been omitted for clarity.  Right panel: Same, but for planet candidates smaller than 4\Rearth.  Red points denote published radii derived from photometry, whilst black points are those published values derived from spectroscopy and blue points are from \textit{Gaia} DR2.  Large error bars arise from uncertainties in the radius ratio $R_p/R_*$ rather than the stellar radii. }
\label{comparison}
\end{figure*}

Figure~\ref{comparison} shows the comparison between planet-candidate radii derived in this work and the values from the literature sources (as per the references given in Table~\ref{tab:planetparams}).  The right panel details planets smaller than 4\,\Rearth\ and differentiates those having previously published radius estimates derived from spectroscopy versus photometry.  We also show results from \citet{kruse19}, who used stellar radii determined from \textit{Gaia} DR2.  No systematic trend is evident in our revised planet radii.  Of the 125 candidates with published spectroscopically-derived radii, for which we obtain $R_p<22$\,\Rearth, our results are 4$\sigma$ different for five of them.  Four of those (EPIC 203070421, 203533312, 210961508, 228754001) orbit evolved stars with log $g$ ranging from 3.31-3.78 and radii from 2.67-3.91\,\Rsun.  This results in larger inferred planetary radii, turning some potentially rocky worlds into gas giants.  Our revised radii for these planet candidates lie in the realm of Saturn and Jupiter, and so remain eminently plausible.

% this is figure 7

\begin{figure}
\includegraphics[width=\columnwidth]{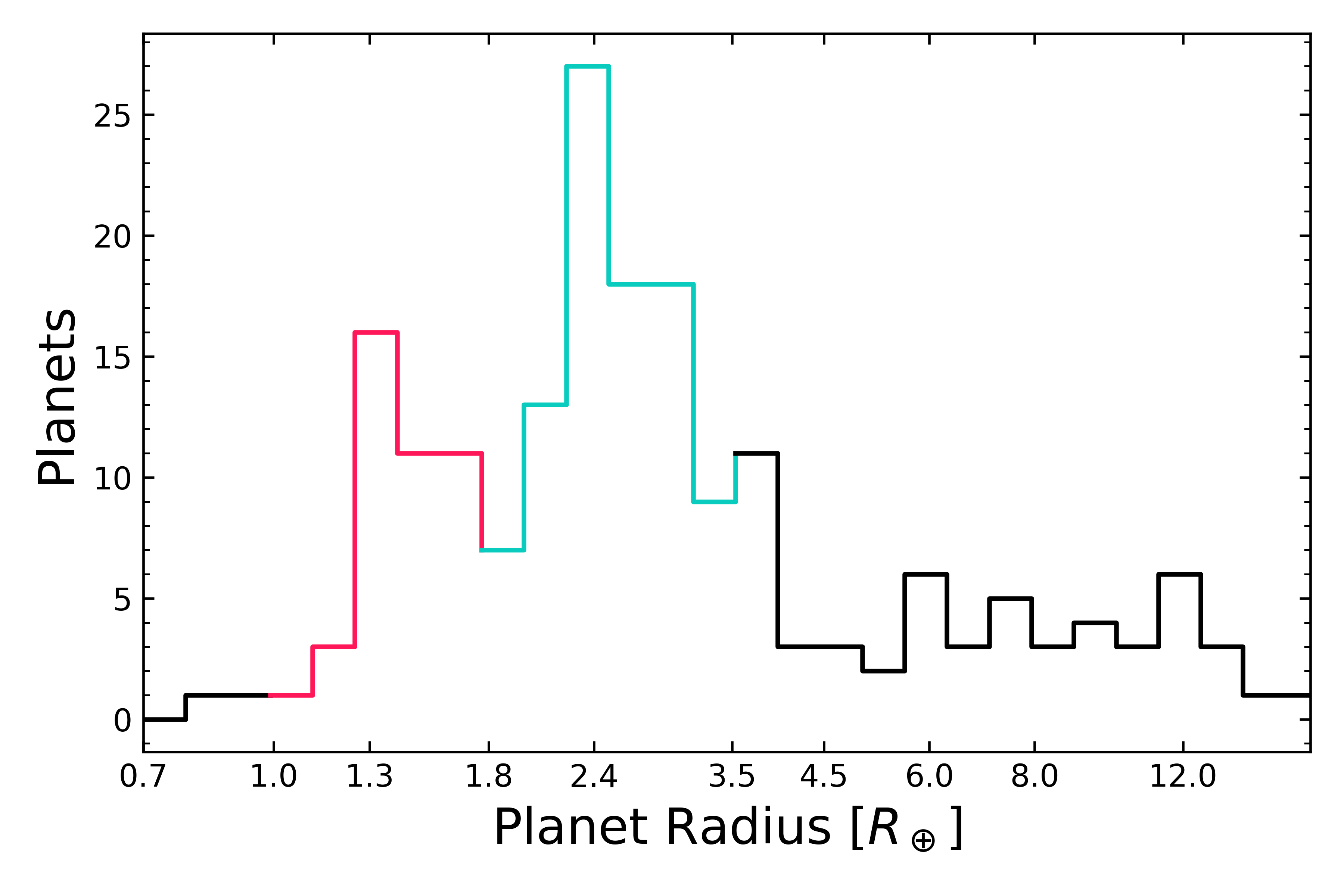}
\caption{Histogram of our revised planet radii. Red: Rocky planets. Cyan: Gaseous ``mini-Neptune'' planets.  The radius gap noted by \citet{fulton17} and \citet{hu20} is evident. }
\label{fig:radiusgap}
\end{figure}

A large-scale analysis of spectroscopic parameters for stars hosting \textit{Kepler} planet candidates revealed a ``radius gap'' \citep{fulton17}, with planets of 1.5-2.0\Rearth\ apparently depleted by more than a factor of two.  Subsequent studies have confirmed that result; \citet{vaneylen18} used 117 planets with median radius uncertainties of 3.3\% as derived from asteroseismology to further characterise the radius gap.  In Figure~\ref{fig:radiusgap}, we show the distribution of planet-candidate radii from our K2-HERMES sample. Our sample, although smaller than the surveys conducted by \citet{fulton17} and \citet{hu20}, also sees a drop off in exoplanetary candidates and confirmed exoplanets centred around 1.8\Rearth.  \citet{hu20} in particular showed that K2 planet candidates were depleted within a radius gap centred at 1.9\,\Rearth.

% If overleaf times out, comment out this figure.. It is the largest. the PDF is even bigger!

% ideally make this a plottwo with the radius gap one first, but that plot breaks Overleaf.
% commented this out since the radius v PERIOD one reproduces the plot in Gupta+19.
%\begin{figure}
%\plotone{evapvalley.pdf}
%\caption{Planet radius versus semimajor axis. The dashed line indicates a hint of the ``evaporation valley'' noted in previous studies of \textit{Kepler} planets \citep[e.g.][]{vaneylen18,gupta19}.  Our median uncertainty in planetary radius is shown as a vertical line. }
%\label{valley}
%\end{figure}

\begin{figure*}
\includegraphics[width=\columnwidth]{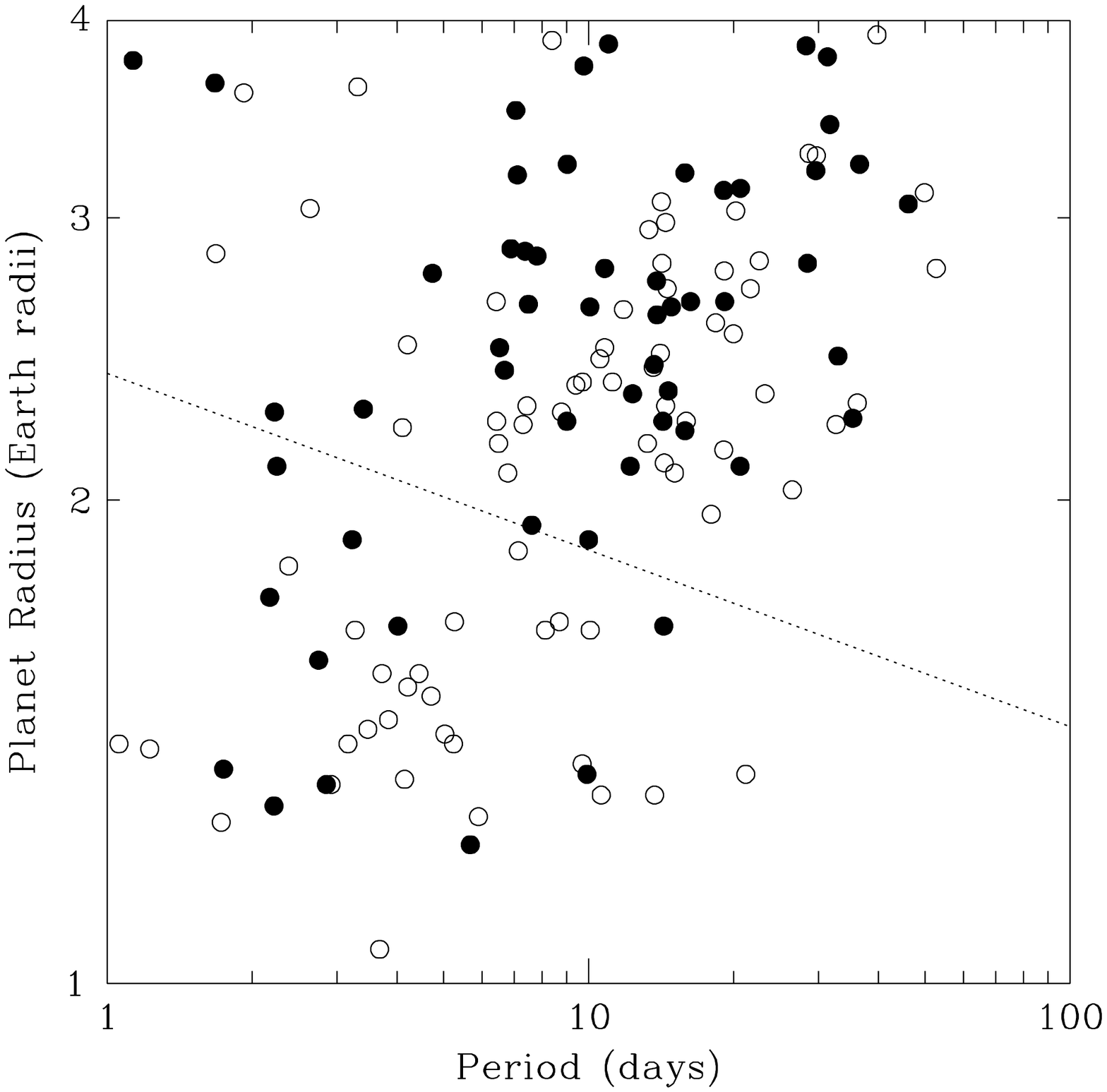}
\includegraphics[width=\columnwidth]{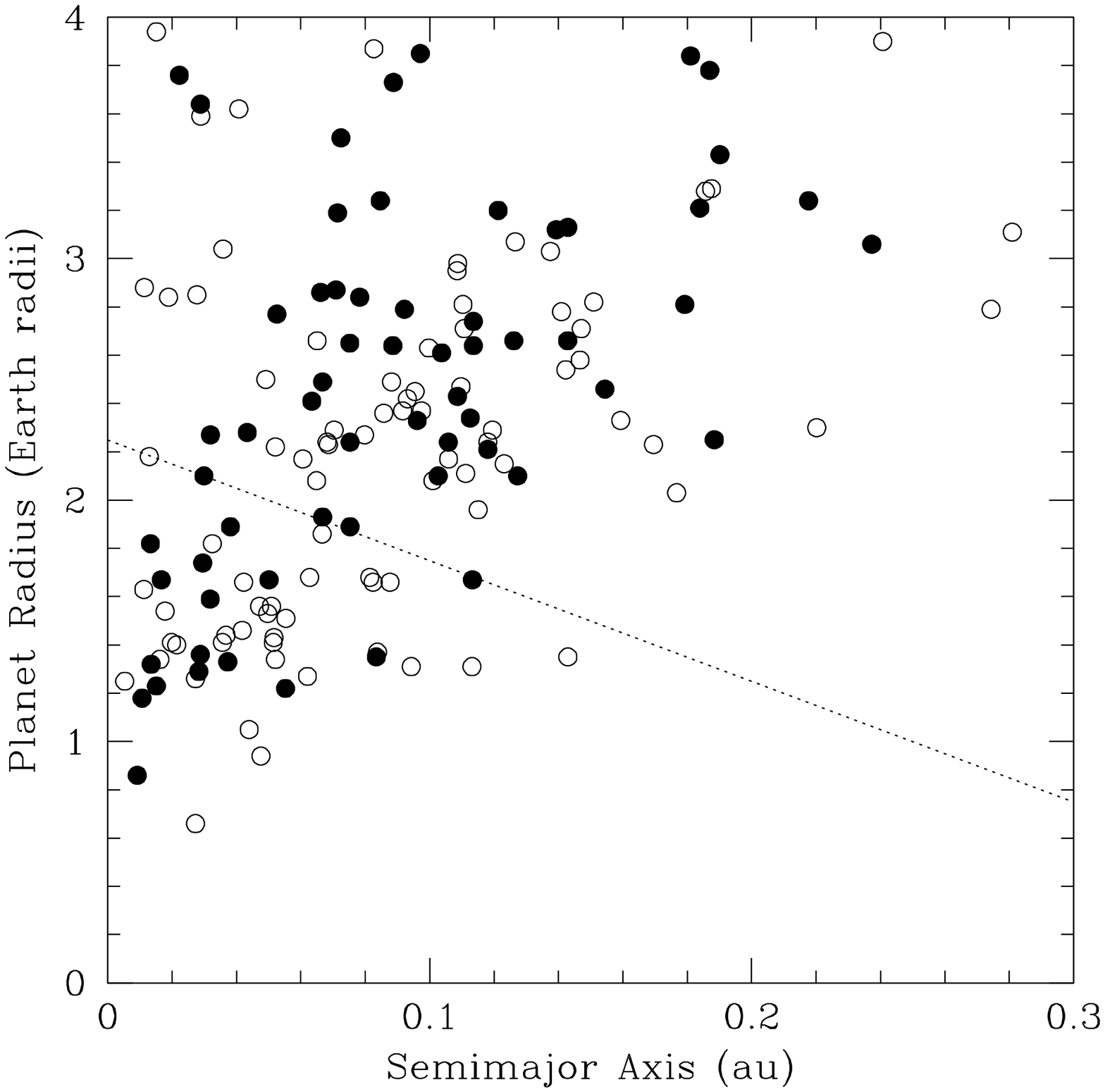}
\caption{Left: Planet radius versus orbital period; filled circles indicate planets for which we obtain radius estimates at better than 10\% precision.  The dashed line indicates the slope in the radius valley as noted by \citet{vaneylen18} and \citet{gupta19}.  Right: Planet radius versus semimajor axis, as computed from the $K2$ period and our derived host-star masses.  Symbols have the same meaning as in the left panel. }
\label{fig:radiivstars2}
\end{figure*}

\begin{figure}
\includegraphics[width=\columnwidth]{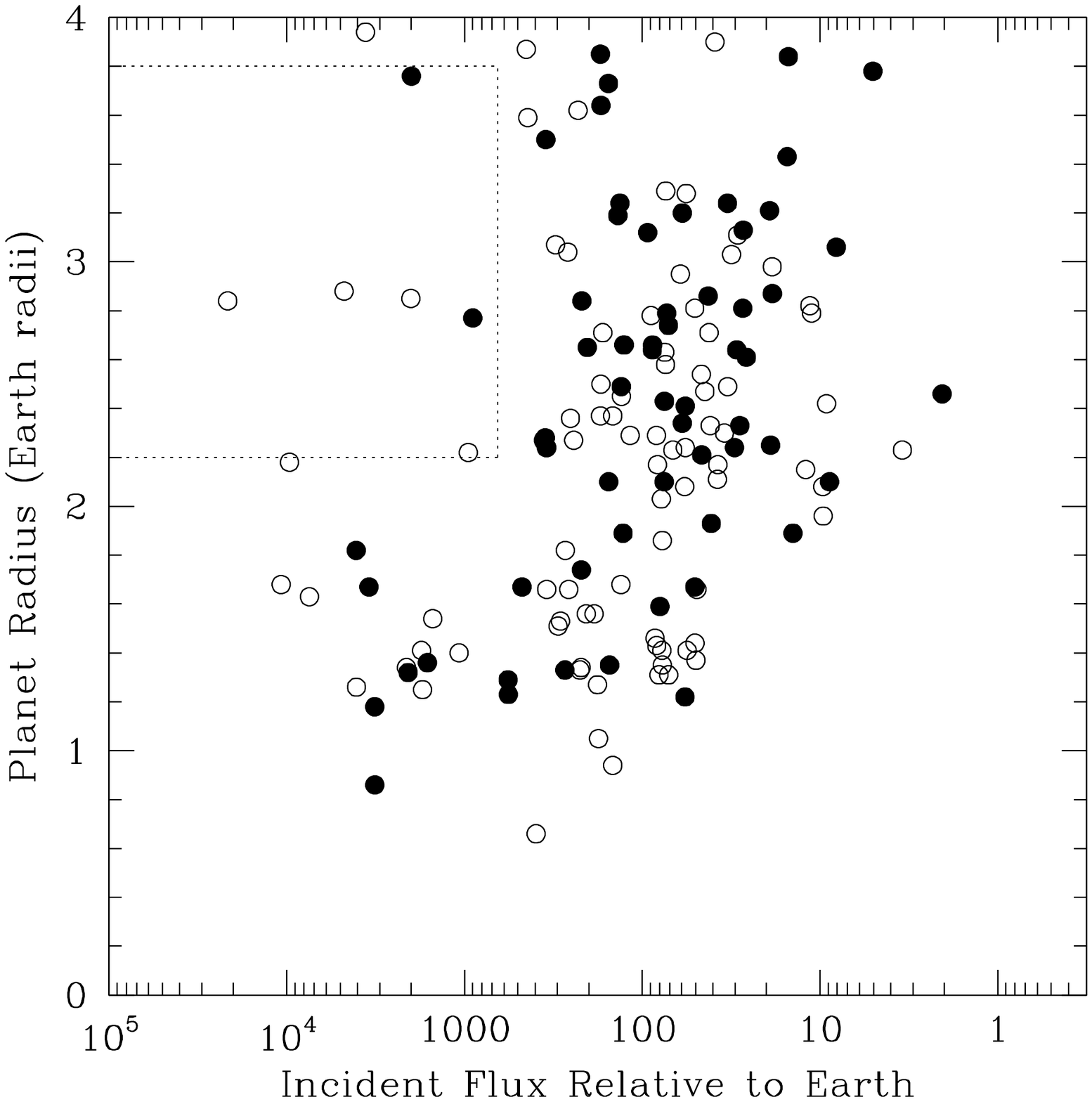}
\caption{Planet radius versus incident flux, in Earth units.  Filled circles indicate planets for which we obtain radius estimates at better than 10\% precision.  The dashed lines enclose the hot Super-Earth desert \citep{lundkvist16}.}
\label{fig:fluxedup}
\end{figure}

In Figure~\ref{fig:radiivstars2}, we explore the radius gap in more detail, showing the planet radii as a function of both orbital period and semimajor axis.  The radius gap was shown by \citet{vaneylen18} to have a slope dependent on orbital period, with a slope of $\frac{{\rm d log} R}{{\rm d log} P}$ of approximately -1/9, a value corroborated by \citet{gupta19} and illustrated in Figure~\ref{fig:radiivstars2}.  In this Figure, we show as filled circles those 95 planets for which we derive radii with precision of 10\% or better.  The K2 sample investigated here gave consistent results for the shape and slope of this evaporation valley, with the exception of four candidates.  These planets (EPIC 206082454, 201754305.02, 210508766.02, 228725791.01) appear as filled circles falling on the dashed line in the right panel of Figure~\ref{fig:radiivstars2}.  These candidates have radii with precisions of better than 10\%.  Interestingly, three of these four are members of multiple systems.  Figure~\ref{fig:fluxedup} gives the planet radius as a function of incident stellar flux (Table~\ref{tab:insolation}).  The hot super-Earth desert postulated by \citet{lundkvist16} is shown as a box enclosing the region between 2.2-3.8\,\Rearth\ and $S_{inc}>$650\,$F_\oplus$.  Near the edges of this region lie only two planet candidates with radius estimates better than 10\% precision, EPIC 206036749.01 and EPIC 211359660.01.

%Figure~\ref{fig:mstar} shows the planet radii as a function of host-star mass.  We see that smaller planets are markedly less prevalent around higher-mass stars, but this is a wholly-expected consequence: stars more massive than 1\,\Msun\ in this sample tend to be late F dwarfs or slightly evolved subgiants, which exhibit higher levels of photometric ``flicker'' \citep{basri13,bastien13,bastien14}, hindering detection of small planets.  The lack of larger planets for stars less than 0.5\,\Msun\ is also consistent with results from radial velocity surveys of M dwarfs \citep[e.g.][]{endl06,hatzes16,tuomi19}.  

%\begin{figure}
%\includegraphics[width=\columnwidth]{radius_v_mstar.pdf}
%\caption{Planet radius versus host-star mass.  Filled circles indicate planets for which we obtain radius estimates at better than 10\% precision. }
%\label{fig:mstar}
%\end{figure}

%  This was figure 9, "has no reference or use in text." Leave out?
%\begin{figure*}
%\includegraphics[width=0.66\columnwidth]{K2HERMESRvTPlot.png}
%\includegraphics[width=0.66\columnwidth]{K2HERMESRvLPlot.png}
%\includegraphics[width=0.66\columnwidth]{K2HERMESRvMPlot.png}
%\caption{Planet radius versus all the stellar parameters.  {\bf Jake: Are these only "CONFIRMED" planets?}  }
%\label{radiivstars1}
%\end{figure*}

%\clearpage
% uncomment this to drop figs before the references.. but then each one gets its own page! 

%-----------------------------------------------------------------------
\section{Summary and Conclusion}

In this work, we have presented a self-consistent catalog of spectroscopic host-star parameters for 199 $K2$ planet hosts, and the derived physical parameters of 224 planets.  We use the revised radii for these planet candidates to cast doubt on 30 as-yet-unconfirmed planets, and we strongly suspect those to be false positives.  We also examine the distribution of planet radii as a function of period, showing that the radius gap of the main \textit{Kepler} sample is indeed also evident in this $K2$ sample.  The slope of the radius valley is also consistent with that obtained for the \textit{Kepler} planets by \citet{vaneylen18} and \citet{gupta19}, with a handful of interesting exceptions. 

In addition to the 30 planet candidates which are rendered implausible based on their revised host-star parameters, our results confirm the small radii of a handful of nearly Earth-sized planets.  They are EPIC\,205470347 (0.66$\pm$0.11\,\Rearth), EPIC\,211357309 (0.86$\pm$0.05\,\Rearth), EPIC\,212464382 (0.94$\pm$0.16\,\Rearth), and EPIC\,220245303 (1.05$\pm$0.18\,\Rearth).  However, as shown in Table~\ref{tab:insolation}, these Earth-size planets are far from Earth-like, receiving stellar flux hundreds of times greater than the Earth.  

%\textbf{probably needs some more ``wrap up'' words here - did we expect trends between planet radius and stellar atmospheric parameters?  Are there any updated planet radii that are surprisingly small or large?}

Our results highlight the importance of accurate stellar parameterisation in the characterisation of newly discovered exoplanets.  Fortunately, with surveys like GALAH and instruments like HERMES it is possible to rapidly characterise large numbers of potential exoplanet host stars.  In the coming decade, as the exoplanet discovery rate continues to climb, such surveys will prove pivotal in ensuring the fidelity of the exoplanet catalogue.

%-----------------------------------------------------------------------
\section*{Acknowledgements}

D.S. is supported by Australian Research Council Future Fellowship FT1400147.  S.S. is funded by University of Sydney Senior Fellowship made possible by the office of the Deputy Vice Chancellor of Research, and partial funding from Bland-Hawthorn's Laureate Fellowship from the Australian Research Council.  S.L.M. acknowledges support from the Australian Research Council through Discovery Project grant DP180101791.  S.B. acknowledges funds from the Alexander von Humboldt Foundation in the framework of the Sofja Kovalevskaja Award endowed by the Federal Ministry of Education and Research. This research has been supported by the Australian Research Council (grants DP150100250 and DP160103747). Parts of this research were supported by the Australian Research Council (ARC) Centre of Excellence for All Sky Astrophysics in 3 Dimensions (ASTRO 3D), through project number CE170100013.  L.C. is supported by Australian Research Council Future Fellowship FT160100402.  This research has made use of NASA's Astrophysics Data System (ADS), and the SIMBAD database, operated at CDS, Strasbourg, France. This research has made use of the NASA Exoplanet Archive, which is operated by the California Institute of Technology, under contract with the National Aeronautics and Space Administration under the Exoplanet Exploration Program. We thank the Australian Time Allocation Committee for their generous allocation of AAT time, which made this work possible.  We acknowledge the traditional owners of the land on which the AAT stands, the Gamilaraay people, and pay our respects to elders past, present, and emerging.

%%%%%%%%%%%%%%%%%%%%%%%%%%%%%%%%%%%%%%%%%%%%%%%%%%

%%%%%%%%%%%%%%%%%%%% REFERENCES %%%%%%%%%%%%%%%%%%

% The best way to enter references is to use BibTeX:

%\bibliographystyle{mnras}
%\bibliography{example} % if your bibtex file is called example.bib

% Alternatively you could enter them by hand, like this:
% This method is tedious and prone to error if you have lots of references

%%%%%%%%%%%%%%%%%%%%%%%%%%%%%%%%%%%%%%%%%%%%%%%%%%

%%%%%%%%%%%%%%%%% APPENDICES %%%%%%%%%%%%%%%%%%%%%

%\appendix

%\section{Some extra material}

%If you want to present additional material which would interrupt the flow of the main paper, it can be placed in an Appendix which appears after the list of references.

%%%%%%%%%%%%%%%%%%%%%%%%%%%%%%%%%%%%%%%%%%%%%%%%%%

% Don't change these lines
\bsp	% typesetting comment
\label{lastpage}
\end{document}